\documentclass[pra,showpacs,amsfonts,amssymb,amsmath,eqsecnum,
tightenlines,nofootinbib,showkeys]{revtex4}
\usepackage{graphicx}

\allowdisplaybreaks

\newcommand{\bra}{\bigl\langle}
\newcommand{\ket}{\bigr\rangle}
\newcommand{\balpha}{\boldsymbol{\alpha}}

\newcommand{\bsigma}{\boldsymbol{\sigma}}
\newcommand{\bSigma}{\boldsymbol{\Sigma}}
\newcommand{\bA}{\boldsymbol{A}}
\newcommand{\bB}{\boldsymbol{B}}
\newcommand{\bE}{\boldsymbol{E}}
\newcommand{\bJ}{\boldsymbol{J}}
\newcommand{\bL}{\boldsymbol{L}}
\newcommand{\bS}{\boldsymbol{S}}
\newcommand{\br}{\boldsymbol{r}}
\newcommand{\bp}{\boldsymbol{p}}
\newcommand{\sA}{\mathbf{A}}
\newcommand{\bH}{\mathbf{H}}
\newcommand{\bQ}{\mathbf{Q}}
\newcommand{\cE}{\mathcal{E}}
\newcommand{\cH}{\mathcal{H}}
\newcommand{\cN}{\mathcal{N}}
\newcommand{\cP}{\mathcal{P}}
\newcommand{\cR}{\mathcal{R}}
\newcommand{\cV}{\mathcal{V}}
\newcommand{\cY}{\mathcal{Y}}
\newcommand{\cW}{\mathcal{W}}
\newcommand{\hA}{\widehat{A}}
\newcommand{\hK}{\widehat{K}}
\newcommand{\hQ}{\widehat{Q}}
\newcommand{\Tr}{\operatorname{Tr}}
\newcommand{\sn}{\operatorname{sn}}
\newcommand{\cn}{\operatorname{cn}}
\newcommand{\dn}{\operatorname{dn}}
\newcommand{\tn}{\operatorname{tn}}
\newcommand{\cosec}{\operatorname{cosec}}
\newcommand{\rmd}{\mathrm{d}}
\newcommand{\rme}{\mathrm{e}}
\newcommand{\rmi}{\mathrm{i}}
\newcommand{\bbC}{\mathbb{C}}
\newcommand{\bbN}{\mathbb{N}}
\newcommand{\bbR}{\mathbb{R}}
\newcommand{\bbZ}{\mathbb{Z}}
\newcommand{\rnu}{\sqrt{\nu}}
\newcommand{\fsl}{\mathfrak{sl}}

\begin{document}


\title{Simultaneous ordinary and type A $\cN$-fold
 supersymmetries\\ in Schr\"odinger, Pauli, and Dirac equations}
\author{Choon-Lin Ho}
\email{hcl@mail.tku.edu.tw}
\author{Toshiaki Tanaka}
\email{ttanaka@mail.tku.edu.tw}
\affiliation{Department of Physics, Tamkang University, Tamsui
25137, Taiwan, R.O.C.}


\begin{abstract}

We investigate physical models which possess simultaneous ordinary
and type A $\cN$-fold supersymmetries, which we call type A
$(\cN,1)$-fold supersymmetry. Inequivalent type A $(\cN,1)$-fold
supersymmetric models with real-valued potentials are completely
classified. Among them, we find that a trigonometric Rosen--Morse
type and its elliptic version are of physical interest. We investigate
various aspects of these models, namely, dynamical breaking and
interrelation between ordinary and $\cN$-fold supersymmetries, shape
invariance, quasi-solvability, and an associated algebra which is
composed of one bosonic and four fermionic operators and dubbed type A
$(\cN,1)$-fold superalgebra. As realistic physical applications, we
demonstrate how these systems can be embedded into Pauli and Dirac
equations in external electromagnetic fields.

\end{abstract}


\maketitle

\section{Introduction}

Ever since the formulation of supersymmetric quantum mechanics
\cite{Wi81}, a much deeper understanding of structure of various
solvable potentials in non-relativistic quantum mechanics has been
gained through the ideas of supersymmetry (SUSY) (for a good review,
see, e.g., \cite{CKS95,Ju96}). For instance, seemingly distinct
potentials, e.g., an infinite square-well and a $\cosec^{2}$
potential, are in fact supersymmetric
partners, and their respective energy levels are connected by
supercharges. The methods of SUSY have also been applied to
obtaining exact spectra of quantum systems with multi-component
wave functions, such as Pauli and Dirac equations. It is based
on the fact that, for some electromagnetic field configurations,
Pauli and Dirac equations possess intrinsic SUSY structure,
and different components of a wave function are related by some
supercharges \cite{AC79,Ui84}. More recently, the method of
SUSY has been employed to construct quasi-exactly solvable
potentials in the Pauli and Dirac equations \cite{HR03,HR04}. For
quasi-exactly solvable systems, it is only possible to determine
algebraically a part of the spectrum but not the whole spectrum
\cite{TU87,Tu88,Sh89,GKO93,Us94}.

Recently, the concepts of SUSY and quasi-exact solvability were
combined within the framework of so-called $\cN$-fold
supersymmetry \cite{AST01b}, which is a natural generalization
\cite{AIS93} of ordinary supersymmetric quantum mechanics. Among
various nonlinear extensions of ordinary SUSY such as
parasupersymmetry \cite{RS88,BD90,Kh93}, fractional supersymmetry
\cite{Du93}, and so on, $\cN$-fold SUSY is characterized by the fact
that anti-commutators of fermionic operators are polynomials of degree
(at most) $\cN$ in bosonic operators. It has been proved in a generic
way that $\cN$-fold SUSY is essentially equivalent to (weak)
quasi-solvability (this latter term, in contrast to quasi-exact
solvability, is used to include the case where the system admits
non-normalizable solutions in closed form) \cite{AST01b}. Up to now,
three different families of $\cN$-fold supersymmetric systems have
been found for arbitrary finite integer $\cN$, namely, type A
\cite{AST01a,Ta03a}, type B \cite{GT03}, and type C \cite{GT04},
which have correspondence with the classification of second-order
linear differential operators preserving a monomial-type vector
space \cite{PT95}.  Of the three types, type A and type C under a
specific condition have been completely classified \cite{Ta03a,GT04}.

While there are still a lot to be done in the mathematical
developments of $\cN$-fold SUSY, it is also interesting that one
looks for physical systems which possess such generalized SUSY. In
view of the fact that so far all the $\cN$-fold SUSY potentials
are only one-dimensional, it is natural that one should look for
physical models which are effectively one-dimensional. Experience
gained in the work in \cite{HR03,HR04} suggests that Pauli and
Dirac equations are good candidates to start with. In this
respect we note that the authors of \cite{Kp01b} found that if the
Pauli equation is generalized such that the gyromagnetic ratio
$g=2$ of the electron is changed to some unphysical values $g=2\cN$
($\cN\geq 2$), then for certain forms of magnetic fields, the
generalized Pauli equation could possess type A $\cN$-fold SUSY.
The result is interesting in a purely mathematical view point, but
unfortunately it would not describe any existing physical systems.

Therefore, we would like to extend the realistic systems in
\cite{HR03,HR04} to include $\cN$-fold SUSY. Since the Pauli and
Dirac equations considered there all possess ordinary SUSY, it is
therefore natural that we look for an ordinary supersymmetric system
which has $\cN$-fold supersymmetry as well, as a starting point of
the aforementioned purpose. This naturally led us further to consider
a more general situation in which a system has simultaneous $\cN$-fold
supersymmetry with two different values of $\cN$. In our previous
paper \cite{HT05a}, we have succeeded in constructing a family of
such systems, which we have called type A $(\cN_{1},\cN_{2})$-fold
SUSY. These systems possess simultaneous type A $\cN_{1}$- and
$\cN_{2}$-fold SUSYs. Hence, the aim mentioned at the beginning of
this paragraph is equivalent to finding Pauli and Dirac systems
with type A $(\cN,1)$-fold SUSY.

In this paper, we shall present a detailed study of Schr\"odinger,
Pauli, and Dirac systems which possess ordinary and type A SUSYs
simultaneously. It is amazing to realize that some well known
shape-invariant potentials, namely, the infinite square-well and
the $\cosec^{2}$ potentials with specific strengths of the coupling
constant, are indeed potentials of such kind. We also find an
elliptic generalization of these potentials which breaks shape
invariance but preserves the structure of type A $(\cN,1)$-fold SUSY.
We study various aspects of these models such as dynamical breaking
and interrelation between ordinary and $\cN$-fold supersymmetries,
shape invariance, quasi-solvability, and an associated algebra which
is composed of one bosonic and four fermionic operators and dubbed
type A $(\cN,1)$-fold superalgebra.

The organization of the paper is as follows. In Sect.~\ref{sec:nsusy}
some of the main ideas of $\cN$-fold SUSY are reviewed. We then
discuss in Sect.~\ref{sec:simul} the idea of simultaneous type A
$(\cN,1)$-fold SUSY.
Section~\ref{sec:schro} gives a detailed discussion of trigonometric
Rosen--Morse type potentials in Schr\"odinger equation. In particular,
we show that shape invariance in the case enables us to establish
a relation between ordinary and $\cN$-fold supercharges, and to derive
type A $(\cN,1)$-fold superalgebra in closed form for arbitrary $\cN$.
Similar discussion of elliptic potentials is given in
Sect.~\ref{sec:ellip}. Extensions to the Pauli and the Dirac
equations are presented in Sect.~\ref{sec:pauli}.
Section~\ref{sec:summa} concludes the paper. In the
Appendix~\ref{sec:class}, we list the complete classification of
real type A $(\cN,1)$-fold SUSY. Appendix~\ref{sec:GBDP} gives a
summary of the generalized Bender--Dunne polynomials (GBDPs), and
present some explicit forms of them for the shape-invariant and
elliptic potentials investigated in Sections \ref{sec:schro} and
\ref{sec:ellip}.

\section{\label{sec:nsusy}$\cN$-fold Supersymmetry}

\subsection{Definition of $\cN$-fold Supersymmetry}

We begin with defining $\cN$-fold supersymmetry in
one-dimensional quantum mechanics. Consider a matrix Hamiltonian
$\bH_{\cN}$ given by
\begin{align}
\bH_{\cN}=\left( \begin{array}{cc} H_{\cN}^{+} & 0\\ 0 & H_{\cN}^{-}
\end{array}\right),
\label{eq:nfham}
\end{align}
where $H_{\cN}^{\pm}$ are ordinary scalar Hamiltonians
\begin{align}
H_{\cN}^{\pm}=\frac{1}{2}p^{2}+V_{\cN}^{\pm}(q),
\label{eq:ordiH}
\end{align}
with $p=-\rmi\rmd /\rmd q$. $\cN$-fold supercharges $\bQ_{\cN}^{\pm}$
are defined by,
\begin{align}
\bQ_{\cN}^{-}= \left( \begin{array}{cc}  0 & P_{\cN}^{-}\\ 0 & 0
\end{array}\right),\qquad
 \bQ_{\cN}^{+}=\left( \begin{array}{cc} 0 & 0\\ P_{\cN}^{+} & 0
\end{array}\right),
\label{eq:dfQ+-}
\end{align}
where the components $P_{\cN}^{\pm}$ are defined by,
\begin{align}
P_{\cN}^{-}=P_{\cN},\qquad P_{\cN}^{+}=P_{\cN}^{t},
\label{eq:dfP+-}
\end{align}
in terms of a monic $\cN$th-order linear differential operator
$P_{\cN}$ of the form,
\begin{align}
P_{\cN}=\frac{\rmd^{\cN}}{\rmd q^{\cN}}+w_{\cN-1}(q)
 \frac{\rmd^{\cN-1}}{\rmd q^{\cN-1}}+\dots+w_{1}(q)
 \frac{\rmd}{\rmd q}+w_{0}(q).
\label{eq:nfsch}
\end{align}
In Eq.~\eqref{eq:dfP+-}, the superscript $t$ denotes the
formal transposition of operators in a linear space of functions
of $q$ defined by $(\rmd/\rmd q)^{t}=-\rmd/\rmd q$. Note that when
all the functions $w_{k}(q)$ appearing in Eq.~\eqref{eq:nfsch}
are real-valued, the operator $P_{\cN}^{+}$ defined by
Eq.~\eqref{eq:dfP+-} is identical with the formal adjoint of
$P_{\cN}$: $P_{\cN}^{+}= P_{\cN}^{\dagger}$.

A system \eqref{eq:nfham} is said to be $\cN$-fold supersymmetric
with respect to $\bQ_{\cN}^{\pm}$ if it commutes with them:
\begin{align}
\bigl[\bQ_{\cN}^{-},\bH_{\cN}\bigr]
 =\bigl[\bQ_{\cN}^{+},\bH_{\cN}\bigr]=0.
\label{eq:dfNfd}
\end{align}
In components, the commutation relations \eqref{eq:dfNfd} are
expressed as the following intertwining relations:
\begin{align}
P_{\cN}^{-}H_{\cN}^{-}-H_{\cN}^{+}P_{\cN}^{-}=0,\qquad
 P_{\cN}^{+}H_{\cN}^{+}-H_{\cN}^{-}P_{\cN}^{+}=0.
\label{eq:inter}
\end{align}
Therefore, the relations in \eqref{eq:inter} give the conditions
for the system $\bH_{\cN}$ to be $\cN$-fold supersymmetric. Note
that the Hamiltonians (\ref{eq:ordiH}) are always symmetric under
the formal transposition even when they are not Hermitian, and
thus each of the relations in Eq.~\eqref{eq:inter} actually
implies the other.

It was investigated in Ref.~\cite{AST01b} that the $\cN$-fold
supersymmetric models defined above have several significant
properties similar to those of the ordinary supersymmetric models.
In the following, we shall summarize some of the most important
aspects of $\cN$-fold supersymmetry.

\subsection{Quasi-solvability, mother Hamiltonian, and generalized
 Witten index}
\label{ssec:QSMHWI}

One of the most important aspects of $\cN$-fold supersymmetry is that
the component Hamiltonians $H^{-}$ and $H^{+}$ are always
\emph{weakly quasi-solvable} with respect to the operators
$P_{\cN}^{-}$ and $P_{\cN}^{+}$, respectively \cite{AST01b,Ta03a}.
That is, $H^{\pm}$ leave the kernels of $P_{\cN}^{\pm}$ invariant:
\begin{align}
H^{\pm}\cV_{\cN}^{\pm}\subset\cV_{\cN}^{\pm},
 \qquad\cV_{\cN}^{\pm}=\ker P_{\cN}^{\pm}.
\end{align}
As a consequence, we can in principle diagonalize algebraically
the Hamiltonians $H^{\pm}$ in the finite $\cN$-dimensional vector
spaces $\cV_{\cN}^{\pm}$, which are thus called \emph{solvable
sectors} of $H^{\pm}$. If the space $\cV_{\cN}^{+(-)}$ is a subspace
of a Hilbert space $L^{2}$ on which the Hamiltonian $H^{+(-)}$ is
defined, the elements of $\cV_{\cN}^{+(-)}$ provide a part of the
exact eigenfunctions and thus $H^{+(-)}$ is called
\emph{quasi-exactly solvable}.

In ordinary SUSY, the anti-commutator of the supercharges
corresponds to the original Hamiltonian $\bH_{\cN}$. However, it is
not generally the case with $\cN$-fold SUSY. This is because $\{
\bQ_{\cN}^{-}, \bQ_{\cN}^{+}\}$ is now a $2\cN$th-order differential
operator. The half of the anti-commutator of $\bQ_{\cN}^{-}$ and
$\bQ_{\cN}^{+}$ is called \textit{mother Hamiltonian} and is denoted
by $\cH_{\cN}$:
\begin{align}
\cH_{\cN}=\frac{1}{2}\bigl\{\bQ_{\cN}^{-},\bQ_{\cN}^{+}\bigr\}
 = \frac{1}{2}\left( \begin{array}{cc} P_{\cN}^{-}P_{\cN}^{+} & 0\\
  0 & P_{\cN}^{+}P_{\cN}^{-}\end{array}\right).
\label{eq:moham}
\end{align}
An immediate consequence of the above definition is that the
mother Hamiltonian always commutes with the $\cN$-fold supercharges,
that is, it is $\cN$-fold supersymmetric:
\begin{align}
\bigl[\bQ_{\cN}^{-},\cH_{\cN}\bigr]=
 \bigl[\bQ_{\cN}^{+},\cH_{\cN}\bigr]=0.
\label{eq:mhrel1}
\end{align}
Furthermore, if the original Hamiltonian $\bH_{\cN}$ is $\cN$-fold
supersymmetric, the mother Hamiltonian also commutes with
$\bH_{\cN}$ due to the relation \eqref{eq:dfNfd}:
\begin{align}
\bigl[\cH_{\cN},\bH_{\cN}\bigr]=0.
\label{eq:cmHmH}
\end{align}

We now come to discuss the characteristics of the spectrum of an
$\cN$-fold supersymmetric system.  As with ordinary SUSY, the
bosonic (lower) and the fermionic (upper) states of such a
system are in one-to-one correspondence provided that these
states are eigenstates of the mother Hamiltonian with
non-zero eigenvalues. This can be seen as follows. Consider a
normalized bosonic state $\Phi^-$  satisfying
\begin{align}
H_{\cN}^{-}\Phi^{-}=E^{-}\Phi^{-}.
\end{align}
Since $\cH_{\cN}$ commutes with $\bH_{\cN}$, $\Phi^{-}$ is also
an eigenstate of $\cH_{\cN}$:
\begin{align}
\cH_{\cN}\begin{pmatrix} 0\\
 \Phi^{-}\end{pmatrix}=\cE\begin{pmatrix} 0\\
 \Phi^{-}\end{pmatrix}.
\end{align}
If $\cE$ is non-zero, then there exists the following
normalized state,
\begin{align}
\Phi^{+}=\frac{1}{\sqrt{2\cE}}P_{\cN}^{-}\Phi^{-}.
\end{align}
{}From Eq.~\eqref{eq:inter}, we have
\begin{align}
H_{\cN}^{+}\Phi^{+}=E^{-}\Phi^{+},
\end{align}
which shows that $\Phi^{+}$ is also an eigenstate of $H_{\cN}^{+}$
with the same eigenvalue $E^{-}$. Furthermore, this state is also
the eigenstate of the mother Hamiltonian with the same $\cE$:
\begin{align}
\cH_{\cN}\begin{pmatrix} \Phi^{+}\\ 0
 \end{pmatrix}=\cE\begin{pmatrix} \Phi^{+}\\ 0
 \end{pmatrix},
\end{align}
since $\cH_{\cN}$ commutes with $\bQ_{\cN}^{+}$ and
\begin{align}
\begin{pmatrix} \Phi^{+}\\ 0
 \end{pmatrix}=\frac{\bQ_{\cN}^{+}}{\sqrt{2\cE}}
 \begin{pmatrix} 0\\ \Phi^{-}
 \end{pmatrix}.
\end{align}
Similarly, one can show that a bosonic state is obtainable
from a fermionic one at each energy level unless $\cE=0$:
\begin{align}
\Phi^{-}=\frac{1}{\sqrt{2\cE}}P_{\cN}^{+}\Phi^{+}.
\end{align}

One can generalize the Witten index of ordinary SUSY to $\cN$-fold
SUSY by
\begin{align}
\Tr (-1)^{F}
 =\dim\ker \bQ_{\cN}^{-}-\dim\ker \bQ_{\cN}^{+},
\end{align}
where $\ker\bQ_{\cN}^{\pm}$ are subspaces of a Hilbert space on
which the super-Hamiltonian $\bH_{\cN}$ is naturally defined.
We note that only the physical states with energy $\cE=0$ of the
mother Hamiltonian $\cH_{\cN}$ contribute to the index. The index
thus takes an integer value as the number of these zero energy states
is finite ($2\cN$ at most). When the generalized Witten index is
non-zero, $\cN$-fold supersymmetry is not broken dynamically.

\subsection{Type A $\cN$-fold supersymmetry}

As previously mentioned, at present three different families of
$\cN$-fold supersymmetric systems have been found for arbitrary
finite integer $\cN$. We shall be concerned only with $\cN$-fold
supersymmetry of type A.

The component of the type A $\cN$-fold supercharge is defined by
\begin{align}
\label{eq:Asc}
P_{\cN}=&\,\biggl(\frac{\rmd}{\rmd q}+W(q)
 -\frac{\cN-1}{2}E(q)\biggr)\biggl(\frac{\rmd}{\rmd q}+W(q)
 -\frac{\cN-3}{2}E(q)\biggr)\times\cdots\notag\\
&\,\cdots\times\biggl(\frac{\rmd}{\rmd q}+W(q)+\frac{\cN-3}{2}
 E(q)\biggr)\biggl(\frac{\rmd}{\rmd q}+W(q)
 +\frac{\cN-1}{2}E(q)\biggr).
\end{align}
According to Ref.~\cite{Ta03a}, the necessary and sufficient
condition for type A $\cN$-fold supersymmetry is the following:
\begin{align}
\label{eq:Apots}
V_{\cN}^{\pm}(q)=\frac{1}{2}W(q)^{2}-\frac{\cN^{\,2}-1}{24}
\bigl(2E'(q)-E(q)^{2}\bigr)\pm\frac{\cN}{2}W'(q)-R,
\end{align}
where $R$ is a constant, and the functions $W(q)$ and $E(q)$
satisfy
\begin{gather}
\label{eq:Acon1} \biggl(\frac{\rmd}{\rmd
q}-E(q)\biggr)\frac{\rmd}{\rmd q}
 \biggl(\frac{\rmd}{\rmd q}+E(q)\biggr)W(q)=0\quad
 \text{for}\quad \cN\ge 2,\\
\biggl(\frac{\rmd}{\rmd q}-2E(q)\biggr)
 \biggl(\frac{\rmd}{\rmd q}-E(q)\biggr)\frac{\rmd}{\rmd q}
 \biggl(\frac{\rmd}{\rmd q}+E(q)\biggr)E(q)=0\quad
 \text{for}\quad \cN\ge 3.
\end{gather}
Construction of potentials possessing type A $\cN$-fold
supersymmetry has been investigated by analytic calculations of
the intertwining relations, and by $\fsl(2)$ construction based on
quasi-solvability.  We refer the readers to \cite{Ta03a} for the
complete classification of the potentials.

The solvable sectors $\cV_{\cN}^{\pm}$ of the type A $\cN$-fold
supersymmetric Hamiltonians, which are the vector spaces preserved
by $H_{\cN}^{\pm}$, are given by the kernel of the type A $\cN$-fold
supercharges:
\begin{align}
\cV_{\cN}^{\pm}=\ker P_{\cN}^{\pm}
 =\rme^{-\cW_\cN^{\pm}}\bra 1,z,\dots,z^{\cN-1}\ket,
\label{sol-space}
\end{align}
where the gauge functions $\cW_{\cN}^{\pm}$ are given by
\begin{align}
\label{eq:gauge}
\cW_{\cN}^{\pm}=\frac{\cN-1}{2}\int\rmd q\,E\mp\int\rmd q\,W.
\end{align}

Finally, we note here that, when $\cN=1$, 1-fold SUSY is just
ordinary SUSY, and $P_{\cN}^{-}$ and $P_{\cN}^{+}$ are simply
ordinary supercharges $A^{-}=\frac{\rmd}{\rmd q}+W$ and
$A^{+}=-\frac{\rmd}{\rmd q}+W$, respectively.

\section{Simultaneous ordinary and $\cN$-fold {SUSY}s}
\label{sec:simul}

Since various quantum systems of physical interest possess
ordinary SUSY, an interesting question is therefore whether some
of these also possess $\cN$-fold SUSY. This question has led us
to investigate a more general problem, namely, to classify
systems which possess
type A $\cN$-fold SUSY with two different values of $\cN$
\cite{HT05a}. We have called such systems as type A
$(\cN_{1},\cN_{2})$-fold supersymmetric.

A system $\bH$ is said to have \emph{$(\cN_{1},\cN_{2})$-fold
supersymmetry} if it commutes with two different $\cN_{i}$-fold
supercharges ($i=1,2$) \emph{simultaneously}, namely,
\begin{align}
\bigl[\bQ_{\cN_{1}}^{(1)\pm},\bH\bigr]
 =\bigl[\bQ_{\cN_{2}}^{(2)\pm},\bH\bigr]=0.
\end{align}
Without loss of generality, we can assume $\cN_{1}\geq\cN_{2}$. In
this case, it is evident that the components $H^{\pm}$ of the system
$\bH$ preserve two vector spaces $\cV_{\cN_{1}}^{(1)\pm}= \ker
P_{\cN_{1}}^{(1)\pm}$ and $\cV_{\cN_{2}}^{(2)\pm}=\ker
P_{\cN_{2}}^{(2)\pm}$ separately, where $P_{\cN_{i}}^{(i)\pm}$ are
components of $\bQ_{\cN_{i}}^{(i)\pm}$ ($i=1,2$).  Hence, the
solvable sectors $\cV_{\cN_{1}\!,\,\cN_{2}}^{\pm}$ of
$(\cN_{1},\cN_{2})$-fold supersymmetric Hamiltonians $H^{\pm}$ are
generally given by
\begin{align}
\cV_{\cN_{1}\!,\,\cN_{2}}^{\pm}=\cV_{\cN_{1}}^{(1)\pm}
 \cup\cV_{\cN_{2}}^{(2)\pm}.
\end{align}

Classification of type A $(\cN_{1},\cN_{2})$-fold supersymmetry
has been presented in \cite{HT05a}.  Here we shall focus on type
A $(\cN,1)$-fold supersymmetric system, i.e., system possessing
simultaneous ordinary and $\cN$-fold SUSYs. For such a system, its
potential must satisfy
\begin{align}
\label{simul}
V_{\cN}^{\pm}(q)&=\frac{1}{2}W(q)^{2}-\frac{\cN^{\,2}-1}{24}
 \bigl(2E'(q)-E(q)^{2}\bigr)\pm\frac{\cN}{2}W'(q)-R \notag\\
&=\frac{1}{2}\left(W_\cN(q)^{2} \pm W_\cN'(q)\right),
\end{align}
where $R$ is a constant, and $W_\cN$ is the first derivative
of a superpotential of ordinary SUSY, with the subscript
$\cN$ signifying that it is also related to $\cN$-fold SUSY.

{}From Eq.~\eqref{eq:Apots} we immediately have
\begin{gather}
\label{eq:cond1}
\cN W'=W'_\cN,\\
\label{eq:cond2}
W^{2}-\frac{\cN^{\,2}-1}{12}(2E'-E^{2})-2R=W_\cN^{2}.
\end{gather}
The first condition \eqref{simul} is easily integrated as
\begin{align}
\label{eq:cond3}
W_\cN=\cN W+C,
\end{align}
where $C$ is a constant. Substituting Eq.~\eqref{eq:cond3} into
Eq.~\eqref{eq:cond2}, we obtain
\begin{align}
\label{eq:cond6}
\left(\cN^{\,2}-1\right) W^{2}+2\cN C W +\frac{\cN^{\,2}-1}{12}
 (2E'-E^{2})+C^{2}+2R=0.
\end{align}
To investigate Eq.~\eqref{eq:cond6}, it is more convenient to
gauge-transform the type A Hamiltonians to \cite{Ta03a}
\begin{align}
\label{eq:gHamA}
H^{\pm}=\rme^{-\cW_\cN^\pm}\biggl[-A(z)\frac{\rmd^{2}}{\rmd z^{2}}
 +\biggl(\frac{\cN-2}{2}A'(z)\pm Q(z)\biggr)\frac{\rmd}{\rmd
 z}
-\frac{(\cN-1)(\cN-2)}{12}A''(z)\pm\frac{\cN-1}{2} Q'(z)-R\biggr]
 \rme^{\cW_{\cN}^{\pm}},
\end{align}
where the gauge potentials $\cW_{\cN}^{\pm}$ are given by
Eq.~\eqref{eq:gauge}, and $A(z)$ and $Q(z)$ are polynomials of at
most fourth- and second-degree, respectively, and related to
$E(q)$ and $W(q)$ by
\begin{align}
\label{eq:defAz}
A(z)&=\frac{1}{2}(z')^{2}=a_{4}z^{4}+a_{3}z^{3}
 +a_{2}z^{2}+a_{1}z+a_{0},\\
\label{eq:defA'z}
A'(z)&=z''=Ez',\\
\label{eq:defQz}
Q(z)&=-Wz'=b_{2}z^{2}+b_{1} z+b_{0}.
\end{align}
With the aid of these relations, we can show that the condition
\eqref{eq:cond6} is satisfied if
\begin{align}
\label{eq:cond8}
\biggl(Q^{2}+\frac{1}{3}H[A]+\frac{2(C^{2}+2R)}{\cN^{\,2}-1}A
 \biggr)^{2}=\frac{8\cN^{\,2}C^{2}}{(\cN^{\,2}-1)^{2}}AQ^{2},
\end{align}
where
\begin{align}
H[A]&= AA''-\frac{3}{4}(A')^{2},
\end{align}
is an algebraic covariant called the Hessian of $A$ \cite{Gu64,Ol99}.
The solvable sectors $\cV_{\cN,\,1}^\pm$ of the type A $(\cN,1)$-fold
supersymmetric Hamiltonians are
\begin{align}
\label{eq:solV}
\cV_{\cN\!,\,1}^{\pm}=\cV_{\cN}^{\pm}\cup\cV_{1}^{\pm},
\end{align}
where $\cV_{\cN}^{\pm}$ are given by Eq.~\eqref{sol-space}, and
$\cV_{1}^{\pm}$ consists of the states
\begin{align}
\psi_0^{\pm}\propto\rme^{\pm Cq}
 \exp\biggl(\pm \cN\int\rmd q\, W\biggr).
\end{align}

We note here that the condition \eqref{eq:cond8} is expressed
as an algebraically covariant form under the projective
transformations $GL(2,K)$ ($K=\bbR$ or $\bbC$) on $A(z)$ and
$Q(z)$ (cf. Refs.\cite{GKO93,Ta03a}):
\begin{align}
\label{eq:transA}
A(z)&\mapsto\hA(z)=\Delta^{-2}(\gamma z+\delta)^{4}
 A\biggl(\frac{\alpha z+\beta}{\gamma z+\delta}\biggr),\\
Q(z)&\mapsto\hQ(z)=\Delta^{-1}(\gamma z+\delta)^{2}
 Q\biggl(\frac{\alpha z+\beta}{\gamma z+\delta}\biggr),
\end{align}
where $\alpha,\beta,\gamma,\delta\in K$ and $\Delta=\alpha\delta
-\beta\gamma\neq0$. We can easily show that the polynomial $A(z)$
of at most fourth-degree is transformed to one of the five
canonical forms listed in Table~\ref{tb:canon} by the $GL(2,\bbC)$
transformation. The complete exhibition of each model for more
general type A $(\cN_{1},\cN_{2})$-fold SUSY in this complex
classification scheme is found in Ref.~\cite{HT05a}.

\begin{table}[ht]
\begin{center}
\begin{tabular}{@{\hspace{10pt}}l@{\hspace{10pt}}l%
@{\hspace{10pt}}l@{\hspace{10pt}}}
\hline
Case & Canonical Form & $H[A]$\\
\hline
I   & $1/2$           & $0$\\
II  & $2z$            & $-3$\\
III & $2\mu z^{2}$    & $-4\mu^{2}z^{2}$\\
IV  & $2\mu(z^{2}-1)$ & $-4\mu^{2}(z^{2}+2)$\\
V   & $2z^{3}-g_{2}z/2-g_{3}/2$
 & $-3z^{3}-3g_{2}z^{2}/2-6g_{3}z-3g_{2}^{2}/16$\\
\hline
\end{tabular}
\end{center}
\caption{Canonical forms of $A(z)$ and their corresponding $H[A]$.
 The parameters $\mu$, $g_{2}$, $g_{3}\in\bbC$ satisfy
 $\mu\neq 0$ and $g_{2}^{3}-27g_{3}^{2}\neq 0$.}
\label{tb:canon}
\end{table}

However, for physical applications where the Hamiltonians should
be real and Hermitian, it is more convenient and suitable to
express potentials in terms of real functions. For this purpose,
we should use the real classification scheme under the
$GL(2,\bbR)$ transformation. In Appendix~\ref{sec:class}, we list
all the inequivalent models of type A $(\cN,1)$-fold SUSY in terms
of real functions. Among these models, we find that only the
Rosen--Morse type and some of the elliptic models can have
physical interest, which we shall discuss in the next two
sections.

\section{Shape-invariant Potential}
\label{sec:schro}

In this section, we concentrate on the trigonometric version of
Case IVa, Eqs.~\eqref{eq:IVa1}--\eqref{eqs:IVa3}. The hyperbolic
version is not of physical interest, as any element of the
solvable sectors $\cV_{\cN}^{\pm}$ and $\psi_{0}^{\pm}$ is
non-normalizable and thus unphysical. For convenience, we shall
hereafter adopt a unit system in which $\hbar=e=2m=c=1$, so that
the results obtained here for the Schr\"odinger equation can be
readily carried over to the Pauli and Dirac equations. As such,
the Hamiltonians $H_{\cN}^{\pm}$ and potentials $V_{\cN}^{\pm}$ in
this and the next two sections differ from those in the previous
sections and in Appendix A by a multiplicative factor 2.

The trigonometric version of Case IVa can be obtained by applying
a scale transformation Eqs.~\eqref{eq:scale1} and \eqref{eqs:scale2}
with $\nu <0$ to Eqs.~\eqref{eq:IVa1}--\eqref{eqs:IVa3}.
We take $\nu =-1/4$ without loss of generality. With this choice,
the functions which characterize the $\cN$-fold ($\cN\geq 2$)
and the ordinary supercharges are given by (we hereafter change
the variable $q$ to $x$)
\begin{align}
\label{eq:EWWN}
E(x)=\cot x,\qquad W(x)=-\frac{1}{2}\cot x,
 \qquad W_{\cN}(x)=-\frac{\cN}{2}\cot x.
\end{align}
The corresponding Hamiltonians read
\begin{align}
H^{\pm}_{\cN}=-\frac{\rmd^{2}}{\rmd x^{2}}+V^{\pm}_{\cN}(x),
 \qquad V^{\pm}_{\cN}(x)=\frac{1}{4}\left[
 \cN(\cN\pm2)\cosec^{2}x -\cN^{\,2}\right].
\label{Hx}
\end{align}
These potentials are periodic in $x$, but tend to positive
infinity at $x=n\pi,~n=0,\pm1,\pm2\ldots$.  Hence, the
particle is confined only in any one of the infinite $\cosec^2$
wells with width $\pi$ in $x$. Without loss of generality, we
shall take the physical domain of interest to be $0<x<\pi$.
The solvable sectors are given by
\begin{align}
\cV_{\cN}^{\pm}=(\sin x)^{-\frac{\cN-1\pm 1}{2}}
 \bra 1,\cos x,\ldots,(\cos x)^{\cN-1}\ket,\qquad
\psi_{0}^{\pm}\propto (\sin x)^{\mp\frac{\cN}{2}}.
\label{sol}
\end{align}
In what follows, we will investigate and discuss various aspects
of simultaneous ordinary and $\cN$-fold supersymmetries,
interplay of them, and their physical consequences.

\subsection{Structure of ordinary {SUSY}}

First of all, due to ordinary SUSY,
the Hamiltonians $H_{\cN}^{\pm}$ are factorizable as
\begin{align}
H_{\cN}^{-}=A_{\cN}^{+}A_{\cN}^{-},\qquad
 H_{\cN}^{+}=A_{\cN}^{-}A_{\cN}^{+},
\label{Hpm}
\end{align}
with the first-order operators
\begin{align}
A_{\cN}^{-}\equiv\frac{\rmd}{\rmd x}+W_{\cN},\qquad
 A_{\cN}^{+}\equiv -\frac{\rmd}{\rmd x}+W_{\cN}.
\label{A_N}
\end{align}
It follows that
\begin{align}
\label{eq:inter2}
A_{\cN}^{-}H_{\cN}^{-}=H_{\cN}^{+}A_{\cN}^{-},\qquad
 A_{\cN}^{+}H_{\cN}^{+}=H_{\cN}^{-}A_{\cN}^{+},
\end{align}
which corresponds to Eq.~\eqref{eq:inter} in the case of $1$-fold
SUSY. If we introduce the supercharges $\sA_{\cN}^{\pm}$ in
a matrix representation by
\begin{align}
\sA_{\cN}^{-}=\left(
 \begin{array}{cc}
 0 & A_{\cN}^{-}\\
 0 & 0
 \end{array}\right),\qquad\sA_{\cN}^{+}=\left(
 \begin{array}{cc}
 0 & 0\\
 A_{\cN}^{+} & 0
 \end{array}\right),
\end{align}
the relations \eqref{Hpm} and \eqref{eq:inter2} are expressed
as a Lie superalgebra:
\begin{align}
\bigl[\sA_{\cN}^{\pm},\bH_{\cN}\bigr]=0,\qquad
 \bigl\{\sA_{\cN}^{-},\sA_{\cN}^{+}\bigr\}=\bH_{\cN}.
\end{align}
Let us denote the normalized eigenfunctions of the
Hamiltonians $H_{\cN}^{\pm}$ by $\psi_{n}^{(\cN)\pm}$ with
eigenvalues $E_{n}^{(\cN)\pm}$, respectively. Here the subscript
$n=0,1,2,\ldots$ denotes the number of nodes of the wave function.
It is easily proved that $V_{\cN}^{-}$ and $V_{\cN}^{+}$, being
partners of ordinary SUSY, have the same energy spectrum except
for the ground state of $V_{\cN}^{-}$ with $E_{0}^{(\cN)-}=0$, which
has no corresponding level for $V_{\cN}^{+}$. More explicitly,
we have the following supersymmetric relations:
\begin{align}
\label{E+E-}
E_{n}^{(\cN)+}&=E_{n+1}^{(\cN)-},\\
\label{wf+}
\psi_{n}^{(\cN)+}&=\bigl(E_{n+1}^{(\cN)-}\bigr)^{-1/2}
 A_{\cN}^{-}\psi_{n+1}^{(\cN)-},\\
\label{wf1}
\psi_{n+1}^{(\cN)-}&=\bigl(E_{n}^{(\cN)+}\bigr)^{-1/2}
 A_{\cN}^{+}\psi_{n}^{(\cN)+}.
\end{align}
Hence $A_{\cN}^{-}$ converts an eigenfunction of $H_{\cN}^{-}$ into
an eigenfunction of $H_{\cN}^{+}$ with the same energy, but with one
less number of nodes, while $A_{\cN}^{+}$ does the reverse.

The supersymmetric element $\psi_{0}^{-}$ of the solvable sectors in
Eq.~\eqref{sol} is normalizable in the physical domain $0<x<\pi$ we
have considered, and corresponds to the ground state wave function
$\psi_{0}^{(\cN)-}$ of $V_{\cN}^{-}$:
\begin{align}
\psi_{0}^{(\cN)-}=\psi_{0}^{-}\propto (\sin x)^{\frac{\cN}{2}},
 \qquad A_{\cN}^{-}\psi_{0}^{(\cN)-}=0.
\label{gdwf}
\end{align}
Hence, ordinary SUSY in this system is unbroken.

\subsection{Two chains of shape-invariant potentials}
\label{ssec:chain}

Next, we note that the potentials $V_{\cN}^{\pm}$ satisfy
\begin{align}
V_{\cN}^{+}(x)=V_{\cN+2}^{-}(x)+R_{\cN},\qquad
 R_{\cN}=\cN+1.
\label{shape}
\end{align}
This implies
\begin{align}
A_{\cN}^{-}A_{\cN}^{+}=A_{\cN +2}^{+}A_{\cN +2}^{-}+R_{\cN},
\label{eq:inter3}
\end{align}
or, equivalently, $H_{\cN}^{+}=H_{\cN +2}^{-}+R_{\cN}$.
Potentials satisfying Eq.~\eqref{shape} are called shape-invariant,
and are (exactly) solvable \cite{CKS95,Ju96}. Indeed, it turns
out that $V_{\cN}^{\pm}$ belong to one of the well known
shape-invariant potentials, namely, the Rosen--Morse-I potentials
\cite{CKS95,Ju96}, only with parameters being related to the
integer $\cN$. In particular, the potential $V_{\cN}^{-}$
corresponding to $\cN=2$ is simply an infinite square-well in
the domain $0<x<\pi$.
Below we give a brief discussion of the relation between shape
invariance and exact solvability, specializing to the potentials
$V_{\cN}^{\pm}$ considered here.

The wave function and energy of the ground state of $V_{\cN+2}^{-}$
are  $\psi_{0}^{(\cN+2)-}\propto (\sin x)^{\frac{\cN+2}{2}}$ (cf.
Eq.~\eqref{gdwf}) and $E_{0}^{(\cN+2)-}=0$, respectively. The shape
invariance \eqref{shape} indicates that the ground state wave
function $\psi_{0}^{(\cN)+}$ of $V_{\cN}^{+}$ is given by
$\psi_{0}^{(\cN)+}=\psi_{0}^{(\cN+2)-}$ with the ground state
energy $E_{0}^{(\cN)+}=R_{\cN}$. On the other hand, from
the supersymmetric relations \eqref{E+E-}--\eqref{wf1},
the first excited state $\psi_{1}^{(\cN)-}$ of $V_{\cN}^{-}$ is
related to the ground state $\psi_{0}^{(\cN)+}$ of $V_{\cN}^{+}$
by $\psi_{1}^{(\cN)-}\propto A_{\cN}^{+}\psi_{0}^{(\cN)+}$, with
the same energy $E_{1}^{(\cN)-}=E_{0}^{(\cN)+}=R_{\cN}$. Thus,
we have $\psi_{1}^{(\cN)-}\propto A_{\cN}^{+}\psi_{0}^{(\cN+2)-}$.
By iterating this process, we can obtain the wave function of
the $n$th excited state of $V_{\cN}^{-}$ as
\begin{align}
\psi_{n}^{(\cN)-}\propto A_{\cN}^{+}A_{\cN+2}^{+}\cdots
 A_{\cN+2(n-2)}^{+}A_{\cN+2(n-1)}^{+}\psi_{0}^{(\cN+2n)-},
\label{wf2}
\end{align}
and energy eigenvalues are given by
\begin{align}
E_{n}^{(\cN)-}&=\sum_{k=0}^{n-1}R_{\cN+2k}
 =n(\cN+n),\qquad n=0,1,2,\ldots.
\label{En}
\end{align}
With these formulas, the energy and wave functions for all
$V_{\cN+2k}^{\pm}$ are also exactly calculable in closed form. We
note here that since Eq.~\eqref{shape} requires the integers $\cN$
in $V_{\cN}^{\pm}$ to be differed by two, we therefore have two
different chains of shape-invariant $\cosec^{2}$ potentials in
which two neighboring potentials possess type A $(\cN,1)$-fold SUSY,
one for even $\cN$'s, and the other for odd $\cN$'s.

To make discussion of $\cN$-fold SUSY structure of $V_{\cN}^{\pm}$
transparent in the next subsection, we find it convenient to
rewrite the symbols of the potentials in the same chain of shape
invariance related by supercharges $A_{\cN+2k}^{-}$ and
$A_{\cN+2k}^{+}$, ($k=0,1,2,\ldots$ and $\cN+2k\geq 2$). Let us write
\begin{align}
\label{Vm}
V^{(\cN+2k)}(x)=V_{\cN+2k}^{-}(x)+\sum_{l=0}^{k-1} R_{\cN+2l},
 \qquad k=0,1,2,\ldots.
\end{align}
The potentials $V^{(m)}$ and $V^{(m+2)}$ possess type A $(m,1)$-fold
SUSY. Now let us denote by $\psi_{n}^{(m)}$  and $E_{n}^{(m)}$
($m=\cN+2k,~n,k=0,1,2,\ldots$) the wave function and energy
eigenvalue, respectively, of the $n$th energy level of the
potential $V^{(m)}$. According to Eq.~\eqref{Vm}, the ground
state of $V^{(m)}$ is also the ground state of $V_{m}^{-}$
(with energy $E_{0}^{(m)-}=0$), hence we have $\psi^{(m)}_{0}\propto
(\sin x)^{\frac{m}{2}}$, and $E^{(m)}_{0}=(m^{2}-\cN^{\,2})/4$ from
Eqs.~\eqref{gdwf} and \eqref{En}. Then the supersymmetric relations
\eqref{Hpm}, \eqref{E+E-}--\eqref{wf1} can be rewritten in terms of
the eigenfunctions $\psi_{n}^{(m)}$ and eigenvalues $E_{n}^{(m)}$
of the potential $V^{(m)}$ defined by Eq.~\eqref{Vm} as follows:
\begin{align}
\label{SUSY1}
A_{m}^{+}A_{m}^{-}\psi_{n}^{(m)}&=E_{n}^{(m)}\psi_{n}^{(m)},\\
\label{SUSY2}
E_{n+1}^{(m)}&=E_{n}^{(m+2)},\\
\label{SUSY3}
A_{m}^{-}\psi_{n}^{(m)}&\propto\psi_{n-1}^{(m+2)},\\
\label{SUSY4}
A_{m-2}^{+}\psi_{n}^{(m)}&\propto\psi_{n+1}^{(m-2)}.
\end{align}
The relations \eqref{SUSY3} and \eqref{SUSY4} are represented
diagrammatically in Fig.~\ref{fig:act} for odd $\cN$.

\subsection{Structure of type A $\cN$-fold {SUSY}}

Now we examine how type A $\cN$-fold SUSY manifests itself in the
two chains of shape-invariant potentials. Essentially three issues
need be addressed, namely, (1) whether $\cN$-fold SUSY is
dynamically broken or not; (2) how the supercharges
$P_{\cN}^{\pm}$ and $A_{\cN}^{\pm}$ are related; and (3) what kind
of superalgebra the five operators $\bH_{\cN},\sA_{\cN}^{\pm},
\bQ_{\cN}^{\pm}$ which characterize the system satisfy. We shall
address these questions below for the two chains of shape-invariant
potentials.

\subsubsection{Dynamical $\cN$-fold {SUSY} breaking}
\label{sssec:broken}

If $\cN$-fold SUSY is not broken, then the ground state, and
perhaps some excited states, must belong to
\begin{align}
\cV_{\cN}^{-}=(\sin x)^{-\frac{\cN}{2}+1}\bra 1,\cos x,\ldots,
 (\cos x)^{\cN-1}\ket.
\label{V-}
\end{align}
In our present case, the ground state of $V_{\cN}^{-}$, given by
Eq.~\eqref{gdwf}, is arranged as
\begin{align}
\psi_{0}^{(\cN)}\propto (\sin x)^{\frac{\cN}{2}}=
 (\sin x)^{-\frac{\cN}{2}+1}(\sin x)^{\cN-1}.
\end{align}
Now it is not hard to check that for odd $\cN$, the factor
$(\sin x)^{\cN-1}$ belongs to $\bra 1,\cos x,\ldots,(\cos x)^{\cN-1}
\ket$, while for even $\cN$ it does not.

Next, we shall consider the first excited states $\psi_{1}^{(\cN)}$.
{}From the relation \eqref{SUSY4}, we find that $\psi_{1}^{(\cN)}$
can be obtained by applying the supercharge $A_{\cN}^{+}$ on
$\psi_{0}^{(\cN+2)}$, i.e.,
\begin{align}
\psi_{1}^{(\cN)}&\propto A_{\cN}^{+}\psi_{0}^{(\cN+2)}
 =\left(-\frac{\rmd}{\rmd x}-\frac{\cN}{2}\cot x\right)
 (\sin x)^{\frac{\cN+2}{2}}\notag\\
&\propto (\sin x)^{\frac{\cN}{2}}\cos x
 =(\sin x)^{-\frac{\cN}{2}+1}(\sin x)^{\cN-1}\cos x.
\end{align}
We easily see that $\psi_{1}^{(\cN)}$ does not belong to
$\cV_{\cN}^{-}$ for any $\cN$.
Similarly, one can check that no higher excited states
belong to $\cV_{\cN}^{-}$, as they pick up higher powers in $\sin$
and $\cos$ terms as $\cN$ increases.

In the same way, one can check that all the physical states of
$V_{\cN}^{+}=V^{(\cN+2)}$ do not belong to $\cV_{\cN}^{+}$
for any $\cN$.
Hence, from the generalized Witten index, we conclude that
$\cN$-fold SUSY is preserved barely by the ground state of
$V_{\cN}^{-}=V_{\cN}^{(\cN)}$ for odd $\cN$, and is completely
broken for even $\cN$.

\subsubsection{Actions of $P_{\cN}^{\pm}$ on eigenstates}
\label{sssec:act}

As we have seen in Eq.~\eqref{eq:EWWN}, we have $E=-2W=\cot x$ in
our case. This significant property enables us to express the
$\cN$-fold supercharges $P_{\cN}^{\pm}$ in terms of the ordinary
supercharges $A_{\cN}^{\pm}$. In fact, the type A $\cN$-fold
supercharge defined by Eq.~\eqref{eq:Asc} in this case becomes
\newpage
\begin{align}
P_{\cN}=\biggl(\frac{\rmd}{\rmd x}
 -\frac{\cN}{2}\cot x\biggr)\biggl(\frac{\rmd}{\rmd x}
 -\frac{\cN-2}{2}\cot x\biggr)
\cdots\biggl(\frac{\rmd}{\rmd x}+\frac{\cN-4}{2}\cot x
 \biggr)\biggl(\frac{\rmd}{\rmd x}+\frac{\cN-2}{2}\cot x\biggr).
\end{align}
Comparing each factor in $P_{\cN}$ to $A_{\cN}^{\pm}$ defined by
Eq.~\eqref{A_N} with $W_{\cN}=-\cN(\cot x)/2$, one has for odd
$\cN=2M+1$ ($M=1,2,3,\ldots$)
\begin{subequations}
\label{P-odd}
\begin{align}
P_{2M+1}^{-}&=(-1)^{M}A_{2M+1}^{-}A_{2M-1}^{-}\cdots
 A_{3}^{-}A_{1}^{-}A_{1}^{+}A_{3}^{+}\cdots A_{2M-3}^{+}
 A_{2M-1}^{+},
\label{P-odda}\\
P_{2M+1}^{+}&=(-1)^{M}A_{2M-1}^{-}A_{2M-3}^{-}\cdots
 A_{3}^{-}A_{1}^{-}A_{1}^{+}A_{3}^{+}\cdots A_{2M-1}^{-}
 A_{2M+1}^{-},
\end{align}
\end{subequations}
and for even $\cN=2M$ ($M=1,2,3,\ldots$)
\begin{subequations}
\label{P-even}
\begin{align}
P_{2M}^{-}&=(-1)^{M}A_{2M}^{-}A_{2M-2}^{-}\cdots
 A_{4}^{-}A_{2}^{-}A_{0}^{+}A_{2}^{+}A_{4}^{+}\cdots A_{2M-4}^{+}
 A_{2M-2}^{+},\\
P_{2M}^{+}&=(-1)^{M}A_{2M-2}^{-}A_{2M-4}^{-}\cdots
 A_{4}^{-}A_{2}^{-}A_{0}^{-}A_{2}^{+}A_{4}^{+}\cdots A_{2M-2}^{+}
 A_{2M}^{+}.
\end{align}
\end{subequations}

\begin{figure}[ht]
\begin{center}
\includegraphics[width=.8\textwidth]{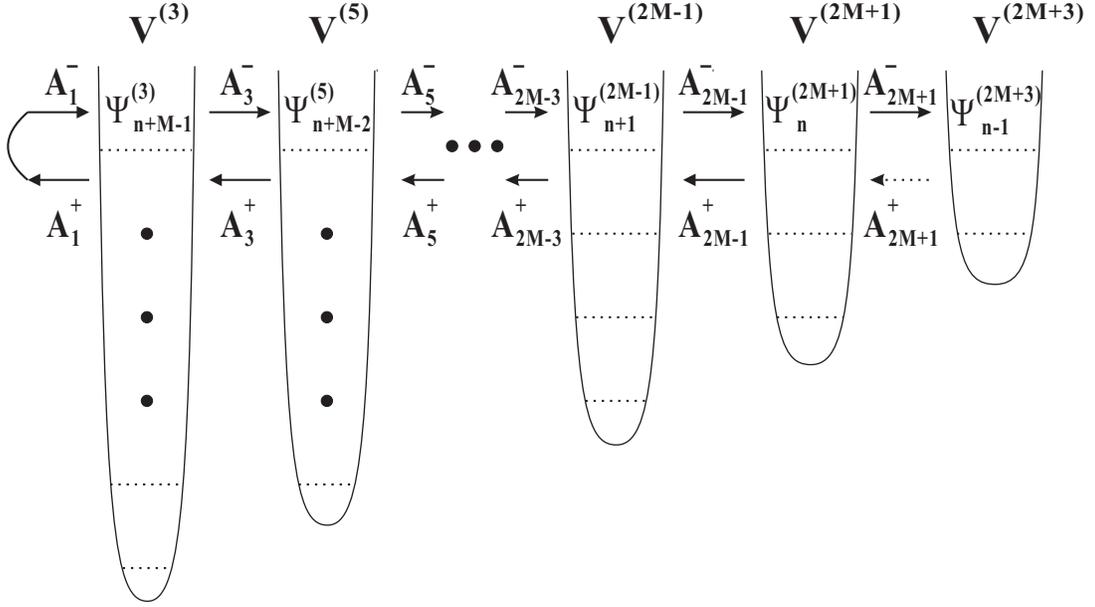}
\caption{Equivalence of the supercharges $P_{2M+1}^{-}$ and $
A_{2M+1}^{-}$ on the state $\psi_{n}^{(2M+1)}$. $P_{2M+1}^{-}$
in Eq.~\eqref{action} is visualized as a series of solid arrows of
$A^{+}$'s (starting with $A_{2M-1}^{+}$) pointing to the left,
turning around with $A_{1}^{-}A_{1}^{+}$, and then pointing to the
right with a series of $A^{-}$'s (ending with $A_{2M+1}^{-}$).
Hence $P_{2M+1}^{-}\psi_{n}^{(2M+1)}\propto A_{2M+1}^{-}
 \psi_{n}^{(2M+1)}\propto\psi_{n-1}^{(2M+3)}$.}
\label{fig:act}
\end{center}
\end{figure}

We first discuss the action of $P_{\cN}^{-}$ on the eigenstates of
$V^{(\cN)}$ for odd $\cN$. As shown in the last section, for odd
$\cN=2M+1$ the ground state $\psi_{0}^{(2M+1)}$, which is
the solution of $A_{2M+1}^{-}\psi_{0}^{(2M+1)}=0$, belongs
to the kernel $\cV_{2M+1}^{-}$ of $P_{2M+1}^{-}$, namely,
$P_{2M+1}^{-}\psi_{0}^{(2M+1)}=0$. Hence, the ground state
$\psi_{0}^{(2M+1)}$ is annihilated by both supercharges
$A_{2M+1}^{-}$ and $P_{2M+1}^{-}$. For the excited states
$\psi_{n}^{(2M+1)}$ ($n>0$) we have
\begin{align}
P_{2M+1}^{-}\psi_{n}^{(2M+1)}\propto A_{2M+1}^{-}(A_{2M-1}^{-}
 A_{2M-3}^{-}\cdots A_{3}^{-})(A_{1}^{-}A_{1}^{+})(A_{3}^{+}
 \cdots A_{2M-3}^{+}A_{2M-1}^{+})\,\psi_{n}^{(2M+1)}.
\label{action}
\end{align}
Applying Eq.~\eqref{SUSY4} repeatedly, we have
\begin{align}
( A_{3}^{+}\cdots A_{2M-3}^{+}A_{2M-1}^{+})\,\psi_{n}^{(2M+1)}
 \propto\psi_{n+M-1}^{(3)},
\end{align}
since each $A^{+}$ maps the $n$th excited state of $V^{(m)}$ to
the $(n+1)$-th state of $V^{(m-2)}$. The action of the $A_{1}^{-}
A_{1}^{+}$ on $\psi_{n+M-1}^{(3)}$ is calculated using
Eqs.~\eqref{eq:inter3} and \eqref{SUSY1} as
\begin{align}
A_{1}^{-}A_{1}^{+}\psi_{n+M-1}^{(3)}
 =(A_{3}^{+}A_{3}^{-}+R_{1})\,\psi_{n+M-1}^{(3)}
 =\left(E^{(3)}_{n+M-1}+R_{1}\right)\,\psi_{n+M-1}^{(3)}.
\end{align}
Hence $A_{1}^{-}A_{1}^{+}\psi_{n+M-1}^{(3)}\propto
\psi_{n+M-1}^{(3)}$. Finally, we note that all the $A^{-}$'s
grouped in the first bracket in the r.h.s. of Eq.~\eqref{action}
simply reverse the actions of the $A^{+}$'s grouped in the third
bracket, and from Eq.~\eqref{SUSY3} we have
\begin{align}
(A_{2M-1}^{-}A_{2M-3}^{-}\cdots A_{3}^{-})\,
 \psi_{n+M-1}^{(3)}\propto\psi_{n}^{(2M+1)}.
\end{align}
Taken together, Eq.~\eqref{action} implies
\begin{align}
P_{2M+1}^{-}\psi_{n}^{(2M+1)}\propto A_{2M+1}^{-}
 \psi_{n}^{(2M+1)}\propto\psi_{n-1}^{(2M+3)}.
\label{P-A}
 \end{align}
Equations \eqref{action} and \eqref{P-A} can be visualized, as shown
in Fig.~\ref{fig:act}, as a series of arrows of $A^{+}$'s (starting
with $A_{2M-1}^{+}$) pointing to the left, turning around with
$A_{1}^{-}A_{1}^{+}$, and then pointing to the right with a series
of $A^{-}$'s (ending with $A_{2M+1}^{-}$).

Hence, the state
$\psi_{n-1}^{(2M+3)}$ is obtainable from the excited state
$\psi_{n}^{(2M+1)}$ ($n>0$) either by the ordinary supercharge
$A_{2M+1}^{-}$, or by the $\cN$-fold supercharge $P_{2M+1}^{-}$.
By the same argument, it can be shown that the actions of
$A_{2M+1}^{+}$ and $P_{2M+1}^{+}$ are equivalent on the state
$\psi_{n-1}^{(2M+3)}$. That all states in the pair of potential
$V^{(\cN)}$ and $V^{(\cN+2)}$ are related by both sets of
supercharges is a manifestation of the underlying unbroken
simultaneous SUSY structures inherent in the $\cosec^{2}$
potential with odd $\cN=2M+1$.

For even $\cN=2M$, type A $\cN$-fold SUSY is broken. Also, the
presence of $A_{0}^{-}$ prevents the $A_{m}^{-}$ ($m=2,4,\ldots,
\cN-2$) to reverse the actions of the corresponding $A_{m}^{+}$ on
$\psi_{n}^{(\cN)}$. Hence, the states in $V^{(\cN)}$ and $V^{(\cN+2)}$
are only related by the supercharges of unbroken ordinary SUSY,
and not by those of broken type A $\cN$-fold SUSY.

\subsubsection{Operator relations among $P_{\cN}^{\pm}$ and
 $A_{\cN}^{\pm}$ for odd $\cN$}

The discussion in the previous subsection indicates that, for odd
$\cN$, the supercharge $P_{\cN}^{\pm}$ are proportional to
$A_{\cN}^{\pm}$ when acting on the eigenstates of $H_{\cN}^{\pm}$,
respectively.
In this section, we will derive the operator relations between
them which show that the $\cN$-fold supercharges $P_{\cN}^{\pm}$
are indeed factorizable into the product of $A_{\cN}^{\pm}$ and
a polynomial in $H_{\cN}^{-}$ or $H_{\cN}^{+}$ for odd $\cN$.

Let us begin with the simplest case. For $\cN=3$, the
supercharges are given by \eqref{P-odda}:
\begin{align}
P_{3}^{-}=-A_{3}^{-}(A_{1}^{-}A_{1}^{+})=-A_{3}^{-}
 (H_{3}^{-}+R_{1})=-(H_{3}^{+}+R_{1})A_{3}^{-},
\label{P3}
\end{align}
where we have made use of the property of shape invariance of
the potentials, namely, Eq.~\eqref{eq:inter3}. This shows that
$P_{3}^{-}$ is factorizable as a product of $A_{3}^{-}$ and a
polynomial in $H_{3}^{\pm}$. The form of $P_{3}^{+}$ is obtained by
taking transposition. Now by repeated use of the shape invariance
relation \eqref{eq:inter3}, one can check that the 5-fold
supercharge $P_{5}^{-}$ is given by
\begin{align}
P_{5}^{-}&=A_{5}^{-}(A_{3}^{-}A_{1}^{-}A_{1}^{+}A_{3}^{+})
 =A_{5}^{-}(H_{5}^{-}+R_{3})(H_{5}^{-}+R_{3}+R_{1})\notag\\
&=(H_{5}^{+}+R_{3})(H_{5}^{+}+R_{3}+R_{1})A_{5}^{-},
\label{P5}
\end{align}
which is again factorizable as a product of $A_{5}^{-}$
and a polynomial in $H_{5}^{\pm}$.

We shall now prove the following formula by induction:
\begin{align}
\label{eq:factP}
P_{2M+1}^{-}&=(-1)^{M}A_{2M+1}^{-}\cdot\prod_{n=0}^{M-1}
 \biggl(H_{2M+1}^{-}+\sum_{k=n}^{M-1}R_{2k+1}\biggr)\notag\\
&=(-1)^{M}\prod_{n=0}^{M-1}\biggl(H_{2M+1}^{+}
 +\sum_{k=n}^{M-1}R_{2k+1}\biggr)\cdot A_{2M+1}^{-}.
\end{align}
Suppose the latter formula holds for a given integer $M$.
{}From Eq.~\eqref{P-odda} and the assumption, we have
\begin{align}
P_{2M+3}^{-}&=-A_{2M+3}^{-}P_{2M+1}^{-}A_{2M+1}^{+}\notag\\
&=(-1)^{M+1}A_{2M+3}^{-}A_{2M+1}^{-}\cdot\prod_{n=0}^{M-1}\biggl(
 H_{2M+1}^{-}+\sum_{k=n}^{M-1}R_{2k+1}\biggr)\cdot A_{2M+1}^{+}.
\end{align}
Using Eqs.~\eqref{Hpm}, \eqref{eq:inter2}, and \eqref{eq:inter3},
we can derive
\begin{align}
P_{2M+3}^{-}&=(-1)^{M+1}A_{2M+3}^{-}A_{2M+1}^{-}A_{2M+1}^{+}
 \cdot\prod_{n=0}^{M-1}\biggl(H_{2M+1}^{+}+\sum_{k=n}^{M-1}
 R_{2k+1}\biggr)\notag\\
&=(-1)^{M+1}A_{2M+3}^{-}(H_{2M+3}^{-}+R_{2M+1})\cdot
 \prod_{n=0}^{M-1}\biggl(H_{2M+3}^{-}+\sum_{k=n}^{M}
 R_{2k+1}\biggr)\notag\\
&=(-1)^{M+1}A_{2M+3}^{-}\cdot\prod_{n=0}^{M}\biggl(H_{2M+3}^{-}
 +\sum_{k=n}^{M}R_{2k+1}\biggr).
\end{align}
The last expression is nothing but the formula \eqref{eq:factP}
with $M$ replaced by $M+1$. Since we have already shown that
Eq.~\eqref{eq:factP} holds for $\cN=3$, Eq.~\eqref{P3}, we
complete the proof of the formula \eqref{eq:factP} for arbitrary
odd integer $\cN=2M+1(\geq 3)$. Equation~\eqref{eq:factP}
naturally explains the equivalent actions of $P_{2M+1}^{-}$ and
$A_{2M+1}^{-}$ on the eigenstates, i.e., Eq.~\eqref{P-A}.

\subsubsection{Type A $(\cN,1)$-fold Superalgebra}

It was proved \cite{AST01b,AS03} that the anti-commutator of
$\cN$-fold supercharges is an $\cN$th-degree polynomial in
the super-Hamiltonian $\bH_{\cN}$, the polynomial being
(proportional to) the characteristic polynomial of the component
Hamiltonians $H_{\cN}^{\pm}$ restricted in the invariant subspaces
$\cV_{\cN}^{\pm}$:
\begin{align}
\bigl\{\bQ_{\cN}^{-},\bQ_{\cN}^{+}\bigr\}\propto\det
 \Bigl(H_{\cN}^{\pm}\bigr|_{\cV_{\cN}^{\pm}}-\bH_{\cN}\Bigr).
\end{align}
However, the direct calculation of the characteristic polynomial
becomes a much harder task as $\cN$ increases. For type A and C
$\cN$-fold supersymmetry, it was shown \cite{Ta03a,GT04} that the
characteristic polynomials are given by the critical generalized
Bender--Dunne polynomials \cite{BD96} and are systematically
calculated through a recursion relation. Hence, in the type A
$(\cN,1)$-fold supersymmetric case, we can also calculate them
systematically using a recursion relation in the same way. We
present some explicit forms of the critical generalized
Bender--Dunne polynomials in Appendix~\ref{sec:GBDP}. Interestingly,
for our present shape-invariant potential, four significant
properties, namely, $\cN$-fold SUSY \eqref{eq:inter}, ordinary SUSY
\eqref{eq:inter2}, shape invariance \eqref{eq:inter3}, and the
relation between ordinary and $\cN$-fold supercharges
\eqref{P-odd}--\eqref{P-even} enable us to calculate directly not
only the anti-commutator of the $\cN$-fold supercharges but also
all the anti-commutators among the ordinary and $\cN$-fold
supercharges.  This will be discussed in what follows.

Let us first begin with the calculation of the anti-commutator
of the $\cN$-fold supercharges:
\begin{align}
\bigl\{\bQ_{\cN}^{-},\bQ_{\cN}^{+}\bigr\}
 =\left( \begin{array}{cc} P_{\cN}^{-}P_{\cN}^{+} & 0\\
 0 & P_{\cN}^{+}P_{\cN}^{-} \end{array}\right),
\end{align}
for odd $\cN=2M+1$. The calculation of the each component in this
case is facilitated by the use of Eq.~\eqref{eq:factP}. In fact,
from Eq.~\eqref{eq:factP}, its transposition and Eq.~\eqref{Hpm}
we have
\begin{align}
P_{2M+1}^{+}P_{2M+1}^{-}
&=\prod_{m=0}^{M-1}\biggl(H_{2M+1}^{-}
 +\sum_{k=m}^{M-1}R_{2k+1}\biggr)\cdot (A_{2M+1}^{+}
 A_{2M+1}^{-})\cdot\prod_{n=0}^{M-1}\biggl(H_{2M+1}^{-}
 +\sum_{l=n}^{M-1}R_{2l+1}\biggr)\notag\\
&=H_{2M+1}^{-}\cdot\prod_{n=0}^{M-1}\biggl(H_{2M+1}^{-}
 +\sum_{k=n}^{M-1}R_{2k+1}\biggr)^{2}\equiv
 \cP_{2M+1}(H_{2M+1}^{-}),
\label{eq:P+P-o}
\end{align}
where $\cP_{2M+1}$ is a monic polynomial of degree $2M+1$. In
a similar way, we can easily show that $P_{2M+1}^{-}P_{2M+1}^{+}
=\cP_{2M+1}(H_{2M+1}^{+})$. Therefore, the anti-commutator for
odd $\cN=2M+1$ reads,
\begin{align}
\bigl\{\bQ_{2M+1}^{-},\bQ_{2M+1}^{+}\bigr\}=\cP_{2M+1}(\bH_{2M+1})
 =\bH_{2M+1}\cdot\prod_{n=0}^{M-1}
 \biggl(\bH_{2M+1}+\sum_{k=n}^{M-1}R_{2k+1}\biggr)^{2}.
\label{eq:Q-Q+o}
\end{align}
For instance, the anti-commutators for $\cN=3$ and $5$ are
\begin{gather}
\bigl\{\bQ_{3}^{-},\bQ_{3}^{+}\bigr\}=\cP_{3}(\bH_{3})
 =\bH_{3}(\bH_{3}+R_{1})^{2},\\
\bigl\{\bQ_{5}^{-},\bQ_{5}^{+}\bigr\}=\cP_{5}(\bH_{5})
 =\bH_{5}(\bH_{5}+R_{3})^{2}(\bH_{5}+R_{3}+R_{1})^{2}.
\end{gather}

For even $\cN=2M$, the supercharge $P_{\cN}^{-}$ ($P_{\cN}^{+}$)
cannot be factorized as a product of $A_{\cN}^{-}$ ($A_{\cN}^{+}$)
and a polynomial in $H_{\cN}^{\pm}$ as in Eq.~\eqref{eq:factP}.
So the previous proof cannot be carried over straightforwardly.
Let us begin with the simplest case, i.e., $\cN=2$. From
Eq.~\eqref{P-even}, we have $P_{2}^{-}=-A_{2}^{-}A_{0}^{+}$.
Hence, $P_{2}^{+}P_{2}^{-}$ can be rewritten, with the use of
the relation $A_{0}^{-}A_{0}^{+}=A_{0}^{+}A_{0}^{-}$ and
Eqs.~\eqref{Hpm} and \eqref{eq:inter3}, as
\begin{align}
\label{eq:P2+P2-}
P_{2}^{+}P_{2}^{-}&=A_{0}^{-}(A_{2}^{+}A_{2}^{-})A_{0}^{+}
 =A_{0}^{-}(A_{0}^{-}A_{0}^{+}-R_{0})A_{0}^{+}\notag\\
&=(A_{0}^{-}A_{0}^{+}-R_{0})A_{0}^{-}A_{0}^{+}
 =H_{2}^{-}(H_{2}^{-}+R_{0}).
\end{align}
Similarly, we can calculate the component $P_{4}^{+}P_{4}^{-}$ as
\begin{align}
P_{4}^{+}P_{4}^{-}
&=A_{2}^{-}A_{0}^{-}A_{2}^{+}A_{4}^{+}A_{4}^{-}A_{2}^{-}
 A_{0}^{+}A_{2}^{+}=A_{2}^{-}A_{0}^{-}(A_{2}^{+}A_{2}^{-}-R_{2})
 A_{2}^{+}A_{2}^{-}A_{0}^{+}A_{2}^{+}\notag\\
&=A_{2}^{-}(A_{0}^{-}A_{0}^{+}-R_{0}-R_{2})(A_{0}^{-}A_{0}^{+}-R_{0})
 A_{0}^{-}A_{0}^{+}A_{2}^{+}\notag\\
&=(A_{2}^{-}A_{2}^{+}-R_{2})A_{2}^{-}
 A_{2}^{+}A_{2}^{-}A_{2}^{+}(A_{2}^{-}A_{2}^{+}+R_{0})\notag\\
&=H_{4}^{-}(H_{4}^{-}+R_{2})^{2}(H_{4}^{-}+R_{2}+R_{0}).
\end{align}

We shall now prove the following formula by induction:
\begin{align}
\label{eq:P+P-e}
P_{2M}^{+}P_{2M}^{-}=H_{2M}^{-}\cdot\prod_{n=1}^{M-1}
 \biggl(H_{2M}^{-}+\sum_{k=n}^{M-1}R_{2k}\biggr)^{2}\cdot
 \biggl(H_{2M}^{-}+\sum_{k=0}^{M-1}R_{2k}\biggr)
 \equiv\cP_{2M}(H_{2M}^{-}),
\end{align}
where $\cP_{2M}$ is a monic polynomial of degree $2M$.
Suppose the latter formula holds for a given integer $M$.
Using the relations \eqref{eq:inter}, \eqref{Hpm}, \eqref{eq:inter3},
and \eqref{P-even}, we have
\begin{align}
P_{2M+2}^{+}P_{2M+2}^{-}
&=A_{2M}^{-}P_{2M}^{+}A_{2M+2}^{+}A_{2M+2}^{-}
 P_{2M}^{-}A_{2M}^{+}\notag\\
&=A_{2M}^{-}P_{2M}^{+}(H_{2M}^{+}-R_{2M})P_{2M}^{-}A_{2M}^{+}\notag\\
&=A_{2M}^{-}P_{2M}^{+}P_{2M}^{-}(H_{2M}^{-}-R_{2M})A_{2M}^{+}.
\end{align}
{}From the assumption and the relations \eqref{Hpm}, \eqref{eq:inter2},
and \eqref{eq:inter3}, we obtain
\begin{align}
P_{2M+2}^{+}P_{2M+2}^{-}
&=A_{2M}^{-}H_{2M}^{-}\cdot\prod_{n=1}^{M-1}\biggl(
 H_{2M}^{-}+\sum_{k=n}^{M-1}R_{2k}\biggr)^{2}\cdot\biggl(
 H_{2M}^{-}+\sum_{k=0}^{M-1}R_{2k}\biggr)(H_{2M}^{-}
 -R_{2M})A_{2M}^{+}\notag\\
&=(H_{2M}^{+})^{2}\cdot\prod_{n=1}^{M-1}\biggl(H_{2M}^{+}
 +\sum_{k=n}^{M-1}R_{2k}\biggr)^{2}\cdot\biggl(H_{2M}^{+}
 +\sum_{k=0}^{M-1}R_{2k}\biggr)(H_{2M}^{+}-R_{2M})\notag\\
&=H_{2M+2}^{-}\cdot\prod_{n=1}^{M}\biggl(H_{2M+2}^{-}
 +\sum_{k=n}^{M}R_{2k}\biggr)^{2}\cdot\biggl(H_{2M+2}^{-}
 +\sum_{k=0}^{M}R_{2k}\biggr).
\end{align}
The last expression is nothing but the formula \eqref{eq:P+P-e}
with $M$ replaced by $M+1$. Since we have already shown that
Eq.~\eqref{eq:P+P-e} holds for $\cN=2$, Eq.~\eqref{eq:P2+P2-},
we complete the proof of the formula \eqref{eq:P+P-e} for
arbitrary even integer $\cN=2M(\geq 2)$. In a similar way, we
can easily show that $P_{2M}^{-}P_{2M}^{+}=\cP_{2M}(H_{2M}^{+})$.
Therefore, for even $\cN=2M$ we have,
\begin{align}
\bigl\{\bQ_{2M}^{-},\bQ_{2M}^{+}\bigr\}=\cP_{2M}(\bH_{2M})
 =\bH_{2M}\cdot\prod_{n=1}^{M-1}\biggl(\bH_{2M}
 +\sum_{k=n}^{M-1}R_{2k}\biggr)^{2}\cdot\biggl(\bH_{2M}
 +\sum_{k=0}^{M-1}R_{2k}\biggr).
\label{eq:Q-Q+e}
\end{align}
For instance, the anti-commutators for $\cN=2$ and $4$ are
\begin{gather}
\bigl\{\bQ_{2}^{-},\bQ_{2}^{+}\bigr\}=\cP_{2}(\bH_{2})
 =\bH_{2}(\bH_{2}+R_{0}),\\
\bigl\{\bQ_{4}^{-},\bQ_{4}^{+}\bigr\}=\cP_{4}(\bH_{4})
 =\bH_{4}(\bH_{4}+R_{2})^{2}(\bH_{4}+R_{2}+R_{0}).
\end{gather}

It is interesting to note that the polynomial system
$\{\cP_{\cN}(E)\}$ for both odd and even $\cN$ satisfies
the following recursion relation with $\cP_{1}(E)\equiv E$ and
$\cP_{0}(E)\equiv 1$:
\begin{align}
\cP_{\cN+2}(E)=E(E+R_{\cN})\,\cP_{\cN}(E+R_{\cN}).
\end{align}

Next, we shall consider the anti-commutators between the ordinary
and $\cN$-fold supercharges:
\begin{align}
\bigl\{\sA_{\cN}^{+},\bQ_{\cN}^{-}\bigr\}=
 \begin{pmatrix}
 P_{\cN}^{-}A_{\cN}^{+} & 0\\
 0 & A_{\cN}^{+}P_{\cN}^{-}
 \end{pmatrix},\qquad
\bigl\{\sA_{\cN}^{-},\bQ_{\cN}^{+}\bigr\}=
 \begin{pmatrix}
 A_{\cN}^{-}P_{\cN}^{+} & 0\\
 0 & P_{\cN}^{+}A_{\cN}^{-}
 \end{pmatrix}.
\end{align}
For odd $\cN=2M+1$, we can easily calculate each component of
the anti-commutators. For instance, using Eqs.~\eqref{Hpm}
and \eqref{eq:factP}, we immediately have
\begin{align}
A_{2M+1}^{+}P_{2M+1}^{-}
&=(-1)^{M}A_{2M+1}^{+}A_{2M+1}^{-}\cdot\prod_{n=0}^{M-1}\biggl(
 H_{2M+1}^{-}+\sum_{k=n}^{M-1}R_{2k+1}\biggr)\notag\\
&=(-1)^{M}\prod_{n=0}^{M}\biggl(H_{2M+1}^{-}+\sum_{k=n}^{M-1}
 R_{2k+1}\biggr)\equiv (-1)^{M}\cR_{M+1}(H_{2M+1}^{-}),
\end{align}
where $\cR_{M+1}$ is a monic polynomial of degree $M+1$ (we shall
adopt the convention that a summation is zero if the upper index
is smaller than the lower one, i.e., $\sum_{n=a}^b f_n=0$ if $b<
a$).  Similarly, we obtain $A_{2M+1}^{-}P_{2M+1}^{+}=(-1)^{M}
\cR_{M+1}(H_{2M+1}^{+})$ and
$P_{2M+1}^{\pm}A_{2M+1}^{\mp}=(-1)^{M}\cR_{M+1}(H_{2M+1}^{\mp})$.
Therefore, the anti-commutators for odd $\cN=2M+1$ reads,
\begin{align}
\label{eq:oddAQ}
\bigl\{\sA_{2M+1}^{+},\bQ_{2M+1}^{-}\bigr\}=\bigl\{\sA_{2M+1}^{-},
 \bQ_{2M+1}^{+}\bigr\}=(-1)^{M}\cR_{M+1}(\bH_{2M+1})
 =(-1)^{M}\prod_{n=0}^{M}\biggl(\bH_{2M+1}+\sum_{k=n}^{M-1}
 R_{2k+1}\biggr).
\end{align}

For even $\cN=2M$, each component in the anti-commutators cannot
be expressed as a polynomial in $H_{\cN}^{-}$ or $H_{\cN}^{+}$.
However, with the aid of Eqs.~\eqref{eq:inter}, \eqref{Hpm},
\eqref{eq:inter2}, and \eqref{eq:Q-Q+e} we have
\begin{subequations}
\begin{align}
A_{\cN}^{+}P_{\cN}^{-}P_{\cN}^{+}A_{\cN}^{-}
&=A_{\cN}^{+}\cdot\cP_{\cN}(H_{\cN}^{+})A_{\cN}^{-}
 =H_{\cN}^{-}\cdot\cP_{\cN}(H_{\cN}^{-}),\\
P_{\cN}^{+}A_{\cN}^{-}A_{\cN}^{+}P_{\cN}^{-}
&=P_{\cN}^{+}H_{\cN}^{+}P_{\cN}^{-}
 =H_{\cN}^{-}\cdot\cP_{\cN}(H_{\cN}^{-}).
\end{align}
\end{subequations}
In a similar way, we can show $P_{\cN}^{-}A_{\cN}^{+}A_{\cN}^{-}
P_{\cN}^{+}=A_{\cN}^{-}P_{\cN}^{+}P_{\cN}^{-}A_{\cN}^{+}
=H_{\cN}^{+}\cdot\cP_{\cN}(H_{\cN}^{+})$. Therefore, combining
the results, we obtain the following algebraic relation:
\begin{align}
\label{eq:AQAQ}
\bigl\{\sA_{\cN}^{+},\bQ_{\cN}^{-}\bigr\}\cdot
 \bigl\{\sA_{\cN}^{-},\bQ_{\cN}^{+}\bigr\}
=\bigl\{\sA_{\cN}^{-},\bQ_{\cN}^{+}\bigr\}\cdot
 \bigl\{\sA_{\cN}^{+},\bQ_{\cN}^{-}\bigr\}
=\bH_{\cN}\cdot\cP_{\cN}(\bH_{\cN}).
\end{align}
In particular, for even $\cN=2M$ we have
\begin{align}
\bigl\{\sA_{2M}^{\pm},\bQ_{2M}^{\mp}\bigr\}\cdot
 \bigl\{\sA_{2M}^{\mp},\bQ_{2M}^{\pm}\bigr\}
=\prod_{n=1}^{M}\biggl(\bH_{2M}+\sum_{k=n}^{M-1}R_{2k}\biggr)^{2}
 \cdot\biggl(\bH_{2M}+\sum_{k=0}^{M-1}R_{2k}\biggr).
\end{align}
The existence of the last factor in the r.h.s. naturally explains
why the anti-commutator of $\sA_{\cN}^{\pm}$ and $\bQ_{\cN}^{\mp}$
for even $\cN$ cannot be expressed as a polynomial in $\bH_{\cN}$.
{}From Eqs.~\eqref{eq:oddAQ} and \eqref{eq:AQAQ}, or directly from
the expressions of $\cP_{2M+1}$ in Eq.~\eqref{eq:P+P-o} and
$\cR_{M+1}$ in Eq.~\eqref{eq:oddAQ}, we see the following relation:
\begin{align}
\cR_{M+1}(E)^{2}=E\,\cP_{2M+1}(E).
\end{align}
Finally, we summarize the complete type A $(\cN,1)$-fold
superalgebra composed of one bosonic and four fermionic operators
$\{\bH_{\cN},\sA_{\cN}^{\pm},\bQ_{\cN}^{\pm}\}$ for both odd and
even $\cN$:\\

\noindent
\emph{Type A $(\cN,1)$-fold superalgebra:}
\begin{subequations}
\label{eqs:AN1alg}
\begin{gather}
\bigl[\sA_{\cN}^{\pm},\bH_{\cN}\bigr]
 =\bigl[\bQ_{\cN}^{\pm},\bH_{\cN}\bigr]=0,\\
\bigl\{\sA_{\cN}^{\pm},\sA_{\cN}^{\pm}\bigr\}
 =\bigl\{\sA_{\cN}^{\pm},\bQ_{\cN}^{\pm}\bigr\}
 =\bigl\{\bQ_{\cN}^{\pm},\bQ_{\cN}^{\pm}\bigr\}=0,\\
\bigl\{\sA_{\cN}^{-},\sA_{\cN}^{+}\bigr\}=\bH_{\cN},\qquad
 \bigl\{\bQ_{\cN}^{-},\bQ_{\cN}^{+}\bigr\}=\cP_{\cN}(\bH_{\cN}),\\
\bigl\{\sA_{\cN}^{\pm},\bQ_{\cN}^{\mp}\bigr\}\cdot\bigl\{
 \sA_{\cN}^{\mp},\bQ_{\cN}^{\pm}\bigr\}=\bH_{\cN}\cdot
 \cP_{\cN}(\bH_{\cN}).
\end{gather}
\end{subequations}
For odd $\cN=2M+1$, the last relation can be decomposed as
\begin{align}
\label{eq:decom}
\bigl\{\sA_{2M+1}^{\pm},\bQ_{2M+1}^{\mp}\bigr\}=
 (-1)^{M}\cR_{M+1}(\bH_{2M+1}).\tag{4.60e}
\end{align}
Using the explicit value of $R_{\cN}$ in Eq.~\eqref{shape},
the monic polynomials $\cP_{\cN}$ and $\cR_{M+1}$ are calculated
as
\begin{align}
\cP_{\cN}(E)&=
 \begin{cases}
 \displaystyle{E\cdot\prod_{n=1}^{M-1}\bigl(E+M^{2}-n^{2}
 \bigr)^{2}\cdot\bigl(E+M^{2}\bigr)} & \text{for\; $\cN=2M$,}\\
 \displaystyle{E\cdot\prod_{n=0}^{M-1}\bigl(E
 +(M-n)(M+n+1)\bigr)^{2}} & \text{for\; $\cN=2M+1$,}
 \end{cases}\label{P(E)}\\
\cR_{M+1}(E)&=\prod_{n=0}^{M}\bigl(E+(M-n)(M+n+1)\bigr).
\end{align}

\section{Elliptic Potential}
\label{sec:ellip}

Next, we consider an example from Case V, i.e., the elliptic case.
For physical systems, it is more relevant to consider this case in
terms of the real Jacobi elliptic functions instead of the
Weierstrass functions discussed in \cite{HT05a}.

A case of physical interest, which corresponds to Case V$'$a
in Appendix~\ref{sec:class}, is given by the potentials
\begin{align}
V_{\cN}^{\pm}(x)=\frac{\cN^{\,2}\cn^{2}x}{4 \sn^{2}x\dn^{2}x}
 \pm\frac{\cN}{2}\left(\frac{1}{\sn^{2}x\dn^{2}x}
 -\frac{k^{2}-k^{\prime 2}}{\dn^{2}x}-1\right),
\label{V}
\end{align}
where $0<k<1$, and $k^{\prime 2}=1-k^{2}$. The functions
in the $\cN$-fold and ordinary supercharges are
\begin{align}
E(x)=\frac{\cn x}{\sn x\dn x}\left(2\dn^{2}x-1\right),
 \qquad W_{\cN}(x)=\cN W(x)=-\frac{\cN\cn x}{2\sn x\dn x}.
\label{E-W}
\end{align}
The solvable sectors are given by
\begin{subequations}
\label{Ker-V5}
\begin{gather}
\cV_{\cN}^{\pm}=(\sn x)^{-\frac{\cN-1\pm 1}{2}}\,(\dn x)^{
 -\frac{\cN-1\mp 1}{2}}\bra 1,\cn x,\ldots,(\cn x)^{\cN-1}\ket,
\label{Ker-V5a}\\
\psi_{0}^{\pm}\propto\left(\frac{\sn x}{\dn x}
 \right)^{\mp\frac{\cN}{2}}.
\label{Ker-V5b}
\end{gather}
\end{subequations}
These potentials are periodic in $x$ with period $2K$, where $K$
is the complete elliptic integral of the first kind satisfying
$\sn K=1$. As in the case of $\cosec^{2}$ potentials, the
particle is confined in one of these periodic intervals bounded by
infinite walls. Hence, we shall take the physical domain to be
$0<x<2K$ without loss of generality. We mention
here that, in the limit $k\to 0$, the Jacobi functions $\sn x$,
$\cn x$, and $\dn x$ reduce to $\sin x$, $\cos x$, and $1$,
respectively. We easily see that, in this limit, the elliptic
system given by Eqs.~\eqref{V}--\eqref{Ker-V5} exactly reduces
to the trigonometric system \eqref{eq:EWWN}--\eqref{sol}
investigated in Sect.~\ref{sec:schro}.
The other limit, $k\to 1$, gives a hyperbolic case, which is
not of physical interest as mentioned previously.
In the following, we shall discuss what kinds of properties remain
unchanged and what kinds of new features arise in comparison with
the trigonometric case in Sect.~\ref{sec:schro}.

\subsection{Structure of ordinary {SUSY}}

Let us begin with the structure of ordinary SUSY. The supersymmetric
relations \eqref{Hpm}--\eqref{wf1} are completely preserved in
the elliptic case with
\begin{align}
A_{\cN}^{\pm}=\mp\frac{\rmd}{\rmd x}-\frac{\cN\cn x}{2\sn x\dn x}.
\end{align}
The supersymmetric element $\psi_{0}^{-}$ of the solvable sectors
in Eq.~\eqref{Ker-V5b} remains normalizable in the physical domain
$0<x<2K$ we have considered, and thus corresponds to the ground
state wave function $\psi_{0}^{(\cN)-}$ of $V_{\cN}^{-}$:
\begin{align}
\label{eq:wf0E}
\psi_{0}^{(\cN)-}=\psi_{0}^{-}\propto\left(\frac{\sn x}{\dn x}
 \right)^{\frac{\cN}{2}},\qquad A_{\cN}^{-}\psi_{0}^{(\cN)-}=0.
\end{align}
Hence, ordinary SUSY is preserved in the elliptic case, too.

Unlike the $\cosec^{2}$-potentials discussed in
Sect.~\ref{sec:schro}, however, the potentials \eqref{V} are not
shape-invariant, as they do not satisfy Eq.~\eqref{shape} with
$x$-independent $R_{\cN}$. Thus, all the remarkable properties of
the trigonometric potentials discussed in Sect.~\ref{ssec:chain}
are lost. In particular, the Hamiltonians with the elliptic
potentials \eqref{V} are only quasi-solvable but are not (exactly)
solvable, which is also guaranteed by the fact $a_{4}\neq 0$ (cf.
Refs.~\cite{Tu88,Sh89}).

\subsection{Structure of $\cN$-fold {SUSY}}

Next, we shall examine the aspect of $\cN$-fold SUSY. The ground
state wave function \eqref{eq:wf0E} can be arranged as
\begin{align}
\psi_{0}^{(\cN)-}\propto (\sn x)^{-\frac{\cN}{2}+1}
 (\dn x)^{-\frac{\cN}{2}}(\sn x)^{\cN-1}.
\end{align}
Thus, as in the trigonometric case, we easily see that for odd
$\cN$ the ground state wave function $\psi_{0}^{(\cN)-}$ is
an element of $\cV_{\cN}^{-}$ given in Eq.~\eqref{Ker-V5a}, but
not for even $\cN$. For the excited states, we cannot proceed
the investigation as in the trigonometric case due to the lack
of shape invariance. However, we can check by using the method of
Bethe ansatz equations \cite{HR03,HR04,Us94} that no elements
of $\cV_{\cN}^{\pm}$ can be normalizable physical eigenstates
except for the ground state $\psi_{0}^{(\cN)-}$ for odd $\cN$.

This means $\cN$-fold SUSY is preserved barely by the ground
state of $V_{\cN}^{-}$ for odd $\cN$ while it is completely
broken for even $\cN$. Therefore, the dynamical aspect of both
ordinary and $\cN$-fold SUSY breaking in the trigonometric model
is unchanged in the elliptic model although some characteristic
features in the former, such as shape invariance and exact
solvability, are lost in the latter.

Type A $(\cN,1)$-fold superalgebra for arbitrary $\cN$ cannot
be obtained explicitly in closed form in the elliptic case.
However, the anti-commutator of the $\cN$-fold supercharges in
this case is still given by the critical GBDP of degree $\cN$ in
$\bH_{\cN}$. In Appendix~\ref{sec:GBDP}, we show the recursion
relation which generates the GBDPs and the first several critical
GBDPs $\pi_{\cN}^{[\cN]}$ for the elliptic model in Case V$'$a.
For the anti-commutators between the ordinary and $\cN$-fold
supercharges, we can derive
\begin{align}
\bigl\{\sA_{\cN}^{\pm},\bQ_{\cN}^{\mp}\bigr\}\cdot\bigl\{
 \sA_{\cN}^{\mp},\bQ_{\cN}^{\pm}\bigr\}=\bH_{\cN}\cdot
 \pi_{\cN}^{[\cN]}(\bH_{\cN}),
\end{align}
using the ordinary and $\cN$-fold SUSY relations \eqref{eq:inter},
\eqref{Hpm}, \eqref{eq:inter2}, and \eqref{eq:antiA}. Hence,
type A $(\cN,1)$-fold superalgebra for the elliptic case is simply
given by Eq.~\eqref{eqs:AN1alg} with $\cP_{\cN}$ generalized to
$\pi_{\cN}^{[\cN]}$ (note that $\pi_{\cN}^{[\cN]}\to\cP_{\cN}$ as
$k\to 0$). Unfortunately, we cannot obtain a closed formula for
the anti-commutators between $\sA_{\cN}^{\pm}$ and $\bQ_{\cN}^{\mp}$
in terms of $\bH_{\cN}$ for both odd and even $\cN$. However, direct
calculations for smaller $\cN$ indicate that the $\cN$-fold
supercharges are factorizable for odd $\cN$ in the elliptic case,
too. Suppose, for instance, $P_{3}^{-}$ is factorizable as
\begin{align}
P_{3}^{-}=-A_{3}^{-}(H_{3}^{-}+R_{c})
 =-A_{3}^{-}(A_{3}^{+}A_{3}^{-}+R_{c}),
\end{align}
where $R_{c}$ is a constant. Substituting Eqs.~\eqref{eq:Asc} and
\eqref{A_N} with $W_{\cN}=\cN W$ in the latter equation, we find
that $E$ and $W$ must satisfy
\begin{gather}
\label{eq:fac31}
2E'-E^{2}+12W^{2}+R_{c}=0,\\
\label{eq:fac32}
E''-2W''+(2W-E)E'+12WW'-WE^{2}+28W^{3}+3R_{c}W=0.
\end{gather}
The first condition \eqref{eq:fac31} is identical with
Eq.~\eqref{eq:cond6} for $\cN=3$ with $C=0$ and $R=R_{c}/3$, and
thus automatically satisfied. Eliminating the constant $R_{c}$
in Eqs.~\eqref{eq:fac31} and \eqref{eq:fac32}, we obtain
\begin{align}
\label{eq:fac33}
E''-2W''-(4W+E)E'+12WW'+2WE^{2}-8W^{3}=0.
\end{align}
In our elliptic case, Case V$'$a, $E$ and $W$ are given by
Eq.~\eqref{E-W}. Using these formulas, we can easily check that
the factorization condition \eqref{eq:fac33} is indeed satisfied.
Another evidence of the factorization comes from the form of the
critical GBDPs for odd $\cN$. From Eq.~\eqref{eqs:cgbdp}, we
easily see that all the critical GBDPs of odd degree (up to $\cN=9$)
have the following form:
\begin{align}
\label{eq:oddpf}
\pi_{2M+1}^{[2M+1]}(E)=E\,\bar{\cR}_{M}(E)^{2},
\end{align}
where $\bar{\cR}_{M}$ is a monic polynomial of degree $M$. The
latter form is in fact a necessary condition for the factorization
of a $(2M+1)$-fold supercharge as $P_{2M+1}^{-}=(-1)^{M}
A_{2M+1}^{-}\bar{\cR}_{M}(H_{2M+1}^{-})$; if the factorization
is the case, we can derive in a similar way of the derivation of
Eq.~\eqref{eq:Q-Q+o}
\begin{align}
\bigl\{\bQ_{2M+1}^{-},\bQ_{2M+1}^{+}\bigr\}=\bH_{2M+1}\cdot
 \bar{\cR}_{M}(\bH_{2M+1})^{2},
\end{align}
and thus Eq.~\eqref{eq:oddpf} follows.
Therefore, we conjecture that a decomposed relation for odd $\cN$
similar to Eq.~\eqref{eq:oddAQ} would hold in the elliptic case,
too (see also the discussion in Section~\ref{sec:summa}).

\section{Pauli and {D}irac Equations}
\label{sec:pauli}

It is well known that, under certain field configurations, the
Pauli and the Dirac equation possess ordinary SUSY structures
\cite{CKS95,Ui84,HR03,HR04}. And this fact has been well made used
of in the studies of exact and quasi-exact solvability of these
equations. The results in the last two sections now allow us to
construct Pauli and Dirac systems with simultaneous ordinary and
$\cN$-fold SUSYs.

\subsection{Pauli equation}

The two-dimensional Pauli equation describes the nonrelativistic
motion of a charged spin $\frac{1}{2}$ particle in an external
magnetic field in a plane. The Hamiltonian is given by
\begin{align}
H=(p_{x}+A_{x})^{2}+(p_{y}+A_{y})^{2}+\frac{g}{2}
 (\nabla\times\bA)_{z}\sigma_{z},
\label{Pauli}
\end{align}
where $p_{x}$ and $p_{y}$ are the momentum operators, $g=2$ is
the gyromagnetic ratio, $\bA$ is the vector potential of the
electromagnetic field, and $\sigma_{z}$ is the Pauli matrix. For
uniform magnetic field, $B_{x}=B_{y}=0$ and $B_{z}=B$, the system
is exactly solvable, giving the well-known Landau levels. On
the other hand, the Aharonov--Casher theorem guarantees that for
any general magnetic field $B_{z}=B(x,y)$ perpendicular to the
$xy$ plane, the ground state is exactly calculable, owing to
the existence of supersymmetry in Eq.~\eqref{Pauli}
\cite{CKS95,AC79}.

Consider a magnetic field in the asymmetric gauge given by the
vector potential
\begin{align}
A_{x}(x,y)=0,\qquad A_{y}(x,y)=-\bar{W}(x),
\end{align}
where $\bar{W}(x)$ is an arbitrary function of $x$. The magnetic
field $\bB$ has components $B_{x}=B_{y}=0$ and $B_{z}=
-\bar{W}'(x)$. The Pauli Hamiltonian is then given by
\begin{align}
H=p_{x}^{2}+\left(p_{y}-\bar{W}(x)\right)^{2}-\bar{W}'(x)\sigma_{z}.
\end{align}
Due to the translational invariance along $y$ direction of
the latter system, $[p_{y},H]=0$, eigenfunctions $\psi$ of $H$
can be factorized as
\begin{align}
\psi (x,y)=\rme^{\rmi k_{y}y}\bar{\psi}(x).
\end{align}
Here $k_{y}$ ($-\infty <k_{y}<\infty$) is an eigenvalue of $p_{y}$,
and $\bar{\psi}(x)$ is a two-component function of $x$. The upper
and lower components of $\bar{\psi}$ are then governed by the
Hamiltonians $H_{-}$ and $H_{+}$ respectively, where
\begin{align}
\label{eq:PHam2}
H=\left(
 \begin{array}{cc}
 H_{-} & 0\\ 0 & H_{+}
 \end{array}
\right),\qquad H_{\mp}=-\frac{\rmd^{2}}{\rmd x^{2}}
 +\left(\bar{W}(x)-k_{y}\right)^{2}\mp\bar{W}'(x).
\end{align}
In this form the SUSY structure of the Pauli equation is manifest,
with $W(x)=\bar{W}(x)-k_{y}$ playing the r\^ole of the first
derivative of a superpotential. Once the upper component of
$\bar{\psi}$ is analytically solved for a nonzero energy,
the lower component can be obtained by applying an appropriate
supercharge on the upper component and vice versa \cite{CKS95,Ju96}.

The SUSY structure of the Pauli equation can be made use of in
several ways. First, from the knowledge of shape-invariant SUSY
potentials, it was found that there are four allowed forms of
shape-invariant $W(x)$ for which the spectrum of the Pauli
equation can be algebraically written down \cite{CKS95}. One of
the four forms gives rise to a uniform magnetic field. Second,
from the close connection of ordinary SUSY with quasi-exact
solvability, several field configurations were constructed giving
rise to quasi-exactly solvable Pauli Hamiltonians which have
underlying $\fsl(2)$ Lie-algebraic structures \cite{HR03}.

Now, we could further construct field configurations so that
the Pauli equation describes a system possessing type A
$(\cN,1)$-fold SUSY. One simply takes $W(x)$ to be
$W_{\cN}(x)=-\cN\cn x/(2\sn x\dn x)$ ($\cN\geq 2$).
The magnetic field is given by
\begin{align}
\label{eq:magne}
B_{z}(x)=-\bar{W}'(x)=\frac{\cN}{2}\biggl(\frac{1}{\sn^{2}x
 \dn^{2}x}+\frac{k^{\prime 2}-k^{2}}{\dn^{2}x}-1\biggr).
\end{align}
Hence, the system \eqref{eq:PHam2} describes a quantum particle
moving along the $y$ direction with the momentum $k_{y}$ and
confined in one of the potential wells in the strips $x\in(2nK,
2(n+1)K)$ ($n\in\bbZ$) formed by the magnetic field
\eqref{eq:magne}. The discussions in Section~\ref{sec:ellip}
tell us that the system is quasi-solvable and only the ground
state with zero energy eigenvalue can be obtained in closed form,
Eq.~\eqref{eq:wf0E}. For the Pauli Hamiltonian, the two-component
ground state wave function $\bar{\psi}_{0}$ in this case is
apparently given by
\begin{align}
\bar{\psi}_{0}(x)=
 \begin{pmatrix}
 \psi_{0}^{(\cN)-}\\ 0
 \end{pmatrix}\propto
 \begin{pmatrix}
 \left(\frac{\sn x}{\dn x}\right)^{\frac{\cN}{2}}\\ 0
 \end{pmatrix}.
\end{align}
All the other two-component functions $\bar{\phi}(x)=(\phi^{-}(x),
\phi^{+}(x))^{t}$ with $\phi^{\mp}(x)\in\cV_{\cN}^{\mp}$,
which are preserved by the Pauli Hamiltonian \eqref{eq:PHam2},
are not normalizable and thus unphysical.

In the limit $k\to 0$, the system reduces to the trigonometric
model in Section~\ref{sec:schro}, which is shape-invariant and
exactly solvable. For the Pauli Hamiltonian \eqref{eq:PHam2} in
this limit, we immediately have
\begin{align}
H\bar{\psi}_{n}(x)=E_{n}^{(\cN)-}\bar{\psi}_{n}(x),\qquad
 \bar{\psi}_{n}(x)=
 \begin{pmatrix}
 \psi_{n}^{(\cN)-}(x)\\ \psi_{n-1}^{(\cN)+}(x)
 \end{pmatrix},\qquad n=0,1,2,\ldots.
\end{align}
where $\psi_{-1}^{(\cN)+}(x)\equiv 0$, the component wave
functions $\psi_{n}^{(\cN)-}$ and $\psi_{n-1}^{(\cN)+}$ are
calculated using Eqs.~\eqref{wf+}, \eqref{gdwf}, and \eqref{wf2},
and the energy eigenvalue $E_{n}^{(\cN)-}$ is given by
Eq.~\eqref{En}. It is interesting to note that if we regard the Pauli
Hamiltonian \eqref{eq:PHam2} as a super-Hamiltonian in the matrix
representation \eqref{eq:nfham}, $\bH_{\cN}=H$, the system
satisfies type A $(\cN,1)$-fold superlagebra \eqref{eqs:AN1alg}
with (note that the upper and lower components of the Hamiltonians
are reversed)
\begin{align}
\sA_{\cN}^{-}=\left(
 \begin{array}{cc}
 0 & 0\\ A_{\cN}^{-} & 0
 \end{array}\right),\qquad\sA_{\cN}^{+}=\left(
 \begin{array}{cc}
 0 & A_{\cN}^{+}\\ 0 & 0
 \end{array}\right),\qquad\bQ_{\cN}^{-}=\left(
 \begin{array}{cc}
 0 & 0\\ P_{\cN}^{-} & 0
 \end{array}\right),\qquad\bQ_{\cN}^{+}=\left(
 \begin{array}{cc}
 0 & P_{\cN}^{+}\\ 0 & 0
 \end{array}\right).
\end{align}
The other cases in Appendix~\ref{sec:class} can be treated in
the same manner, which results in different configurations of
magnetic field.

\subsection{Dirac equations}

Dirac equation which couples minimally to a stationary vector
potential can be treated as in the Pauli equation described in the
last section, since the square of the Dirac Hamiltonian in this
form is proportional to the Pauli equation up to an additive
constant. We shall therefore not repeat the discussions. Instead,
we will give an example of Dirac system coupled non-minimally to
an electric field, which is described by the Dirac--Pauli equation
\cite{HR04}.

Consider the motion of a neutral fermion of spin $\frac{1}{2}$ with
mass $m$ and an anomalous magnetic moment $\mu$, in an external
static electric field $\bE$. The fermion is described by a
time-independent four-component spinor $\psi$ which obeys the
Dirac--Pauli equation
\begin{align}
\label{eq:DiPa}
H\psi=\cE\psi,\qquad H=\balpha\cdot\bp
 +\rmi\mu\beta\balpha\cdot\bE+\beta m,
\end{align}
where $\bp=-\rmi\nabla$ and the Dirac matrices are given,
in the standard representation, by
\begin{align}
\balpha = \left( \begin{array}{cc} 0 & \bsigma\\ \bsigma & 0
\end{array}\right),\qquad
\beta= \left( \begin{array}{cc} 1 & 0\\ 0 & -1
\end{array}\right),
\end{align}
where $\bsigma$ are the Pauli matrices. We introduce two-component
spinors $\chi$ and $\varphi$ by $\psi=(\chi, \varphi)^{t}$.
Then the Dirac--Pauli equation \eqref{eq:DiPa} becomes
\begin{subequations}
\label{H1}
\begin{align}
\bsigma\cdot(\bp-\rmi\mu\bE)\chi &=(\cE+m)\varphi,\\
\bsigma\cdot(\bp+\rmi\mu\bE)\varphi &=(\cE-m)\chi.
\end{align}
\end{subequations}

When the electric field is radial, $\bE=E_{r}(r)\hat{\br}$, we have
a set of commuting operators $\{H,\bJ^{2},J_{z},\bS^{2}=3/4,\hK\}$
where $\bJ$ is the total angular momentum $\bJ =\bL+\bS$, $\bL$ is
the orbital angular momentum, $\bS=\frac{1}{2}\bSigma$ is the spin
operator, and $\hK$ is defined as $\hK=\beta(\bSigma\cdot\bL+1)$.
Simultaneous eigenfunctions of these commuting operators can be
written as
\begin{align}
\psi=\begin{pmatrix}\chi\\ \phi\end{pmatrix}=\frac{1}{r}
 \begin{pmatrix} f_{-}(r)\cY^{\kappa}_{jm_{j}}(\theta,\phi)\\
 \rmi f_{+}(r)\cY^{-\kappa}_{jm_{j}}(\theta,\phi)\end{pmatrix},
\end{align}
where $\cY^{\kappa}_{jm_{j}}(\theta,\phi)$ is the spin harmonics
satisfying
\begin{align}
\bJ^{2}\cY^{\kappa}_{jm_{j}}&=j(j+1)\cY^{\kappa}_{jm_{j}},\qquad
 j=\frac{1}{2},\,\frac{3}{2},\ldots,\\
J_{z}\cY^{\kappa}_{jm_{j}}&= m_{j}\cY^{\kappa}_{jm_{j}},\qquad
 |m_{j}|\leq j,\\
\hK\cY^{\kappa}_{jm_{j}}&=-\kappa\,\cY^{\kappa}_{jm_{j}},\qquad
 \kappa=\pm\Bigl(j+\frac{1}{2}\Bigr),\\
(\bsigma\cdot\hat{\br})\cY^{\kappa}_{jm_{j}}
 &=-\cY^{-\kappa}_{jm_{j}}.
\end{align}
With these relations, Eq.~\eqref{H1} reduces to
\begin{subequations}
\label{eqs:f+-}
\begin{align}
\left(\frac{\rmd}{\rmd r}+\frac{\kappa}{r}+\mu E_{r}(r)\right)
 f_{-}(r)&=\left(\cE+m\right)f_{+}(r),
\label{f-}\\
\left(-\frac{\rmd}{\rmd r}+\frac{\kappa}{r}+\mu E_{r}(r)\right)
 f_{+}(r)&=\left(\cE-m\right)f_{-}(r).
\label{f+}
\end{align}
\end{subequations}
This shows that $f_{-}$ and $f_{+}$ form a one-dimensional SUSY
pair, with the first derivative of a superpotential being
\begin{align}
W(r)=\frac{\kappa}{r}+\mu E_{r}(r).
\label{W1}
\end{align}

Now if we choose $W\equiv W_{\cN}=-\cN\cn r/(2\sn r\dn r)$
($\cN\geq 2$), the Dirac--Pauli equation \eqref{eqs:f+-} becomes
a system possessing type A $(\cN,1)$-fold SUSY in its lower and
upper components, just like the situation in the Pauli equation
discussed before. The required electric field configuration is
given by
\begin{align}
\label{eq:Ef}
\mu E_{r}(r)=-\frac{\kappa}{r}-\frac{\cN\cn r}{2\sn r\dn r}.
\end{align}
In this case, the system \eqref{eqs:f+-} describes a Dirac
particle confined in a sphere $r<2K$ or between two spherical
surfaces $2nK<r<2(n+1)K$ ($n\in\bbN$) by the electric field
\eqref{eq:Ef}. From Eq.~\eqref{eqs:f+-}, the components $f_{-}$
and $f_{+}$ satisfy
\begin{align}
H_{\cN}^{\mp}f_{\mp}(r)=A_{\cN}^{\pm}A_{\cN}^{\mp}f_{\mp}(r)
 =(\cE^{2}-m^{2})f_{\mp}(r),
\end{align}
where $A_{\cN}^{\pm}$ are defined as Eq.~\eqref{A_N} with $x$
replaced by $r$. Hence, the lowest eigenfunction $\psi_{0,\kappa}$
with energy eigenvalue $\cE_{0,\kappa}=m$ for the given integer
value of $\kappa$ is exactly computed as
\begin{align}
\label{eq:wf0k}
\psi_{0,\kappa}=\frac{1}{r}
 \begin{pmatrix}\psi_{0}^{(\cN)-}(r)\cY^{\kappa}_{jm_{j}}
 (\theta,\phi)\\ 0
 \end{pmatrix},
\end{align}
where $\psi_{0}^{(\cN)-}$ is defined by Eq.~\eqref{gdwf}.
Due to $\cN$-fold supersymmetry, the system also admits local
analytic solutions with
\begin{align}
f_{\mp}(r)\in\cV_{\cN}^{\mp}=(\sn r)^{\frac{\cN-1\mp 1}{2}}
 (\dn r)^{\frac{\cN-1\pm 1}{2}}\bra 1,\cn r,\ldots,
 (\cn r)^{\cN-1}\ket.
\end{align}
As discussed in Section~\ref{sssec:broken}, they are not
normalizable except for the configuration of $\psi_{0,\kappa}$ in
Eq.~\eqref{eq:wf0k}.

In the limit $k\to 0$, the Dirac--Pauli system \eqref{eqs:f+-}
gets shape invariance discussed in Section~\ref{ssec:chain}. Owing
to it, infinite number of eigenfunctions $\psi_{n,\kappa}$ and
corresponding eigenvalues $\cE_{n,\kappa}$ ($n=0,1,2,\ldots$) for
the given integer value of $\kappa$ can be computed exactly as
\begin{align}
\psi_{n,\kappa}=\frac{1}{r}\begin{pmatrix}
 \psi_{n}^{(\cN)-}(r)\cY^{\kappa}_{jm_{j}}(\theta,\phi)\\
 \rmi\psi_{n-1}^{(\cN)+}(r)\cY^{-\kappa}_{jm_{j}}(\theta,\phi)
 \end{pmatrix},\qquad \cE_{n,\kappa}=\sqrt{m^{2}+E_{n}^{(\cN)-}},
\end{align}
where $\psi_{-1}^{(\cN)+}(r)\equiv 0$, $\psi_{n}^{(\cN)-}$ and
$\psi_{n-1}^{(\cN)+}$ are obtained from Eqs.~\eqref{wf+},
\eqref{gdwf}, and \eqref{wf2}, and $E_{n}^{(\cN)-}$ is given
by Eq.~\eqref{En}.

The other cases in Appendix~\ref{sec:class} can be treated in
the same manner, which results in different configurations of
electric field.

\section{Discussion and Summary}
\label{sec:summa}

In this paper, we have discussed the interesting issue of
simultaneous ordinary and type A $\cN$-fold SUSYs, which we have
called type A $(\cN,1)$-fold SUSY. For realistic systems with
real-valued potentials, there are essentially eight inequivalent
type A $(\cN,1)$-fold supersymmetric models: one is conformal (Case
II), three of them are hyperbolic (trigonometric) including the
Rosen--Morse type (Cases IVa, IVb, IV$'$a), and the other four
cases are elliptic (Cases Va, Vb, V$'$a, V$''$a). Of these models,
the trigonometric Rosen--Morse type and some of the elliptic models
turn out to be of physical interest.

We have fully investigated the trigonometric Rosen--Morse type,
namely, $\cosec^{2}$ potentials and found that when they have
type A $(\cN,1)$-fold SUSY, they constitute two chains of
shape-invariant potentials, one is the even $\cN$ chain and
the other is the odd $\cN$ chain. The even $\cN$ chain starts with
the well-known SUSY pair of a square-well and a Rosen--Morse
potentials in \cite{CKS95}. For both of the chains, ordinary
SUSY is always unbroken. On the other hand, $\cN$-fold SUSY is
completely broken for the even $\cN$ chain while it is unbroken
barely by the ground state for the odd $\cN$ chain.
We have further shown that type A $(\cN,1)$-fold SUSY together
with shape invariance enable us to obtain the complete type A
$(\cN,1)$-fold superalgebra, which is composed of one bosonic
and four fermionic operators $\{\bH_{\cN},\sA_{\cN}^{\pm},
\bQ_{\cN}^{\pm}\}$, in closed form for arbitrary $\cN$.

For the elliptic models, we have examined the one which reduces
to the $\cosec^{2}$ potential in the limit $k\to 0$. We have found
that the dynamical aspects of both ordinary and $\cN$-fold SUSYs
are completely the same as the trigonometric case although shape
invariance and exact solvability are lost in the elliptic case.
We have also shown that the structure of type A $(\cN,1)$-fold
superalgebra is essentially preserved and that it is obtained
through a recursion relation though we cannot obtain a closed
formula for arbitrary $\cN$.

As physical applications, we have presented how type A
$(\cN,1)$-fold SUSY can be embedded in the Pauli and Dirac
systems, namely, a nonrelativistic spin $\frac{1}{2}$ charged
particle in an external static magnetic field in 2 space dimensions
and a Dirac particle coupled non-minimally to an external static
electric field in 3 space dimensions. Some consequences of the
symmetries have been also discussed briefly.

Relation to some mathematical theorems in Ref.~\cite{AS03} also
gets clearer. Applying the theorem on (in)dependence of
supercharges to $(\cN,1)$-fold SUSY, for instance, the necessary
and sufficient condition for $\cN$-fold supercharges
$P_{\cN}^{\pm}$ to be factorizable as a product of a polynomial in
Hamiltonians $H_{\cN}^{\pm}$ and ordinary supercharges
$A_{\cN}^{\pm}$ is
\begin{align}
\label{eq:faccn}
\bigl\{\sA_{\cN}^{+},\bQ_{\cN}^{-}\bigr\}
 =\bigl\{\sA_{\cN}^{-},\bQ_{\cN}^{+}\bigr\}.
\end{align}
In our present cases, the latter condition holds for the
shape-invariant potentials in Section~\ref{sec:schro} with odd
$\cN$ (cf. Eq.~\eqref{eq:oddAQ}),
and the $\cN$-fold supercharges are indeed factorized as
Eq.~\eqref{eq:factP} in the case. On the other hand, for the
shape-invariant potentials with even $\cN$, we have from
Eqs.~\eqref{eq:inter2}, \eqref{P-even}, and $A_{0}^{+}=-A_{0}^{-}$
\begin{align}
\bigl\{\sA_{2M}^{+},\bQ_{2M}^{-}\bigr\}
 =-\bigl\{\sA_{2M}^{-},\bQ_{2M}^{+}\bigr\},
\end{align}
and thus the condition \eqref{eq:faccn} is not satisfied. It could
be possible that the $\cN$-fold supercharges for even $\cN$ are
factorizable as a product of a polynomial in the Hamiltonians and
a differential operator of even order $2k$ ($0<k<M$) which has no
factor proportional to $A_{\cN}^{\pm}$. But direct calculations
for smaller even $\cN$ indicate that it would not be the case. In
other words, $A_{\cN}^{\pm}$ and $P_{\cN}^{\pm}$ for even $\cN$
would be an \emph{optimal} set \cite{AS03}. For the elliptic
models in Section~\ref{sec:ellip}, although we have not been able
to (dis)prove the relation \eqref{eq:faccn} for arbitrary odd and
even $\cN$, direct calculations for smaller $\cN$ also indicate
that factorizability of the $\cN$-fold supercharges would hold for
arbitrary odd $\cN$.

To the best of our knowledge, it is the first time that both
ordinary and $\cN$-fold SUSYs are discussed not only in
Schr\"odinger equations but also in Pauli and Dirac equations.
Hence, we are expecting that further studies in this direction
would reveal much more novel properties and rich structures in
various physical systems. In what follows, we would like to
mention a couple of interesting future issues as examples.

Recently, the so-called $\mathcal{PT}$-symmetric quantum
theories~\cite{BB98} have attracted much attention. Applications
of type A $(\cN_{1},\cN_{2})$-fold SUSY to
$\mathcal{PT}$-symmetric theories would be straightforward since
the framework of $\cN$-fold SUSY does not rely on the concepts of
(pseudo-)Hermiticity and so on. For this aim, one should simply
employ another suitable classification scheme different from the
real classification in this paper.

The present approaches to embed $\cN$-fold SUSY into the Pauli and
Dirac systems essentially rely on the intrinsic structure of
ordinary SUSY. On the other hand, there exists a Dirac system
which has no intrinsic SUSY structure but is quasi-(exactly)
solvable \cite{CH02}. One of the peculiar aspects of the system is
that the underlying algebraic structure is the Lie superalgebra
$\mathfrak{osp}(2|2)$ instead of $\fsl(2)$ \cite{CH05}. Hence, it
is interesting to clarify whether or not we are able to embed
$\cN$-fold SUSY into such a system, too.

\begin{acknowledgments}
This work was supported in part by the National Science Council of
the Republic of China through Grant No. NSC 93-2112-M-032-009.
\end{acknowledgments}

\appendix

\section{Classification of real type A $(\cN,1)$-fold supersymmetric
 models}
\label{sec:class}

In this appendix, we shall present a detailed classification of
\emph{real} type A $(\cN,1)$-fold supersymmetric models. For this
purpose, we must use the classification scheme of real polynomial
of fourth-degree under the real projective transformation
$GL(2,\bbR)$. In Table~\ref{tb:realc}, we show the real canonical
forms of $A(z)$ (second column) and the complex projective
transformations which convert the real canonical form to the
corresponding complex one in Table~\ref{tb:canon} (third column).

\begin{table}[ht]
\begin{center}
\begin{tabular}{@{\hspace{10pt}}l@{\hspace{10pt}}l%
@{\hspace{10pt}}l@{\hspace{10pt}}}
\hline
Case & Real Canonical Form & Transformations\\
\hline
I     & $1/2$               & Identity\\
II    & $2z$                & Identity\\
III   & $\pm 2\nu z^{2}$    & Identity with $\mu=\pm\nu$\\[5pt]
IV    & $\pm 2\nu(z^{2}-1)$ & Identity with $\mu=\pm\nu$\\
IV$'$ & $\pm 2\nu(z^{2}+1)$ & $\beta=\gamma=0$,
 $\delta=\rmi\alpha$ with $\mu=\pm\nu$\\[5pt]
V     & $\pm\frac{\nu}{2}(1-z^{2})(1-k^{2}z^{2})$
 & $\gamma=\alpha$, $\delta=(4k^{2}-k^{\prime 2})\beta/
 (4+k^{\prime 2})$\\
V$'$  & $\pm\frac{\nu}{2}(1-z^{2})(k^{\prime 2}+k^{2}z^{2})$
 & $\gamma=\alpha$, $\delta=-(1+4k^{2})\beta/(1+4k^{\prime 2})$\\
V$''$ & $\pm\frac{\nu}{2}(1+z^{2})(1+k^{\prime 2}z^{2})$
 & $\gamma=\rmi\alpha$, $\delta=\rmi(4k^{\prime 2}-k^{2})\beta/
 (4+k^{2})$\\
\hline
\end{tabular}
\end{center}
\vspace{10pt}
\caption{Real canonical forms of $A(z)$ and complex projective
 transformations which convert them to the corresponding complex
 ones in Table~\ref{tb:canon}. The parameters $\nu$, $k$, $k'\in\bbR$
 satisfy $\nu > 0$, $0<k<1$ and $k^{2}+k^{\prime 2}=1$.}
\label{tb:realc}
\end{table}

We note that the choice of the canonical forms is not unique,
and we select them so that the change of variable determined by
Eq.~\eqref{eq:defAz} is as simple as possible; this is justified
by the fact that the final form of the potentials does not depend
on the choice of the canonical forms, cf. Ref.~\cite{Ta03a},
Section 4. As a consequence, our canonical forms are different
from, e.g., those in Ref.~\cite{Gu64}.

Furthermore, we note that by Eq.~\eqref{eq:defAz} a rescaling of
the parameters $a_{i}$, $b_{i}$, $R$ in the models by an overall
nonzero constant factor $\nu$ has the following effect on the
change of variable:
\begin{align}
\label{eq:scale1}
z(q;\nu a_{i},\nu b_{i},\nu R)=z(\rnu q;a_{i},b_{i},R).
\end{align}
{}From this equation and Eqs.~\eqref{eq:gauge}, \eqref{eq:gHamA},
\eqref{eq:defA'z}, and \eqref{eq:defQz}, we easily obtain the
scaling relations:
\begin{subequations}
\label{eqs:scale2}
\begin{align}
E(q;\nu a_{i},\nu b_{i},\nu R)&=\rnu E(\rnu q;a_{i},b_{i},R),\\
W(q;\nu a_{i},\nu b_{i},\nu R)&=\rnu\, W(\rnu q;a_{i},b_{i},R),\\
\cW(q;\nu a_{i},\nu b_{i},\nu R)&=\cW(\rnu q;a_{i},b_{i},R),\\
V^{\pm}(q;\nu a_{i},\nu b_{i},\nu R)
 &=\nu V^{\pm}(\rnu q;a_{i},b_{i},R).
\end{align}
\end{subequations}
Therefore, we can set $\nu=1$ in the canonical forms III--V$''$
without loss of generality; the models corresponding to an
arbitrary value of $\nu$ follow easily from Eqs.~\eqref{eq:scale1}
and \eqref{eqs:scale2}.

In all the cases, the condition \eqref{eq:cond8} can be
satisfied if and only if $C=0$.
In Case IV, the condition \eqref{eq:cond8} requires a further
separate analysis depending on whether $b_{0}=0$ or $b_{1}=0$.
Similarly, in the real elliptic cases, namely, Cases V--V$''$, we
must make a further separate analysis depending on
whether $b_{2}=b_{0}=0$ or $b_{1}=0$. Some of them however have
no real solutions of Eq.~\eqref{eq:cond8}, for which we shall
omit the presentation of the models. We also neglect Cases I and
III since both cases lead to a trivial model, as has been already
shown in Ref.~\cite{HT05a}.

\subsection{Case II: $A(z)=2z$, $z(q)=q^{2}$}
\label{ssec:case2}

\noindent
\emph{Parameters:}
\begin{align}
b_{2}=b_{1}=0,\qquad b_{0}=1,\qquad R=0.
\end{align}
\emph{Supercharge:}
\begin{align}
E(q)=\frac{1}{q},\qquad W(q)=-\frac{1}{2q}.
\end{align}
\emph{Potentials:}
\begin{align}
V_{\cN\!,1}^{\pm}(q)=\frac{\cN(\cN\pm 2)}{8q^{2}}.
\end{align}
\emph{Solvable sectors:}
\begin{align}
\cV_{\cN}^{(A)\pm}=q^{-\frac{\cN-1\pm 1}{2}}
 \bra 1,q^{2},\dots,q^{2(\cN-1)}\ket,\qquad
 \psi_{0}^{\pm}(q)\propto q^{\mp\frac{\cN}{2}}.
\end{align}

\subsection{Case IV: $A(z)=2(z^{2}-1)$, $z(q)=\cosh 2q$}
\label{ssec:case4}

\subsubsection{Case IVa: $b_{0}=0$}

\noindent
\emph{Parameters:}
\begin{align}
b_{2}=b_{0}=0,\qquad b_{1}=2,\qquad R=-\frac{\cN^{\,2}-1}{3}.
\end{align}
\emph{Supercharge:}
\begin{align}
\label{eq:IVa1}
E(q)=\frac{2\cosh 2q}{\sinh 2q},\qquad
 W(q)=-\frac{\cosh 2q}{\sinh 2q}.
\end{align}
\emph{Potentials:}
\begin{align}
\label{eq:IVa2}
V_{\cN\!,1}^{\pm}=\frac{\cN(\cN\pm 2)}{2\sinh^{2}2q}
 +\frac{\cN^{\,2}}{2}.
\end{align}
\emph{Solvable sectors:}
\begin{subequations}
\label{eqs:IVa3}
\begin{gather}
\cV_{\cN}^{(A)\pm}=(\sinh 2q)^{-\frac{\cN-1\pm 1}{2}}
 \bra 1,\cosh 2q,\dots,(\cosh 2q)^{\cN-1}\ket,\\
 \psi_{0}^{\pm}(q)\propto(\sinh 2q)^{\mp\frac{\cN}{2}}.
\end{gather}
\end{subequations}

\subsubsection{Case IVb: $b_{1}=0$}

\noindent
\emph{Parameters:}
\begin{align}
b_{2}=b_{1}=0,\qquad b_{0}=2,\qquad R=\frac{\cN^{\,2}-1}{6}.
\end{align}
\emph{Supercharge:}
\begin{align}
E(q)=\frac{2\cosh 2q}{\sinh 2q},\qquad W(q)=-\frac{1}{\sinh 2q}.
\end{align}
\emph{Potentials:}
\begin{align}
V_{\cN\!,1}^{\pm}=\frac{\cN(\cN\mp 2)}{
 2\sinh^{2}2q}\pm\frac{\cN}{2\sinh^{2}q}.
\end{align}
\emph{Solvable sectors:}
\begin{subequations}
\begin{gather}
\cV_{\cN}^{(A)\pm}=(\sinh 2q)^{-\frac{\cN-1}{2}}
 (\tanh q)^{\mp\frac{1}{2}}\bra 1,\cosh 2q,\dots,
 (\cosh 2q)^{\cN-1}\ket,\\
\psi_{0}^{\pm}(q)\propto (\tanh q)^{\mp\frac{\cN}{2}}.
\end{gather}
\end{subequations}

\subsection{Case IV$'$: $A(z)=2(z^{2}+1)$, $z(q)=\sinh 2q$}
\label{ssec:case4'}

\subsubsection{Case IV$\,'\!$a: $b_{0}=0$}

\noindent
\emph{Parameters:}
\begin{align}
b_{2}=b_{0}=0,\qquad b_{1}=2,\qquad R=\frac{\cN^{\,2}-1}{3}.
\end{align}
\emph{Supercharge:}
\begin{align}
E(q)=\frac{2\sinh 2q}{\cosh 2q},\qquad
 W(q)=-\frac{\sinh 2q}{\cosh 2q}.
\end{align}
\emph{Potentials:}
\begin{align}
V_{\cN\!,1}^{\pm}=\frac{\cN(\cN\pm 2)}{
 2\cosh^{2}2q}-\frac{\cN^{\,2}}{2}.
\end{align}
\emph{Solvable sectors:}
\begin{subequations}
\begin{gather}
\cV_{\cN}^{(A)\pm}=(\cosh 2q)^{-\frac{\cN-1\pm 1}{2}}
 \bra 1,\sinh 2q,\dots,(\sinh 2q)^{\cN-1}\ket,\\
\psi_{0}^{\pm}(q)\propto(\cosh 2q)^{\mp\frac{\cN}{2}}.
\end{gather}
\end{subequations}

\subsubsection{Case IV$\,'\!$b: $b_{1}=0$}

In this case, there are no real solutions of Eq.~\eqref{eq:cond8}
for $b_{0}$.

\subsection{Case V: $A(z)=\frac{1}{2}(1-z^{2})(1-k^{2}z^{2})$,
 $z(q)=\sn q$}
\label{ssec:case5}

\subsubsection{Case Va: $b_{2}=b_{0}=0$}

\noindent
\emph{Parameters:}
\begin{align}
b_{2}=b_{0}=0,\qquad b_{1}=\frac{k^{\prime 2}}{2},\qquad
 R=\frac{\cN^{\,2}-1}{12}(1+k^{2}).
\end{align}
\emph{Supercharge:}
\begin{align}
E(q)=-\frac{\sn q(2\dn^{2}q-k^{\prime 2})}{\cn q\dn q},\qquad
 W(q)=-\frac{k^{\prime 2}}{2}\frac{\sn q}{\cn q\dn q}.
\end{align}
\emph{Potentials:}
\begin{align}
V_{\cN\!,1}^{\pm}=\frac{\cN^{\,2}k^{\prime 4}\sn^{2}
 q}{8\cn^{2}q\dn^{2}q}\mp\frac{\cN k^{\prime 2}}{4}\biggl(
 \frac{1+k^{2}}{\dn^{2}q}+\frac{k^{\prime 2}}{\cn^{2}q\dn^{2}q}
 -1\biggr).
\end{align}
\emph{Solvable sectors:}
\begin{subequations}
\begin{gather}
\cV_{\cN}^{(A)\pm}=(\cn q)^{-\frac{\cN-1\mp 1}{2}}
 (\dn q)^{-\frac{\cN-1\pm 1}{2}}\bra 1,\sn q,\dots,
 (\sn q)^{\cN-1}\ket,\\
\psi_{0}^{\pm}\propto\biggl(
 \frac{\dn q}{\cn q}\biggr)^{\mp\frac{\cN}{2}}.
\end{gather}
\end{subequations}

\subsubsection{Case Vb: $b_{1}=0$}

In this case, we have two sets of real solutions of
Eq.~\eqref{eq:cond8}, namely, $b_{2}=\frac{k}{2}(1\pm k)$
and $b_{0}=-\frac{1}{2}(k\pm 1)$. Here we only show the result
corresponding to the upper-sign solutions.\\

\noindent
\emph{Parameters:}
\begin{align}
b_{2}=\frac{k(1+k)}{2},\qquad b_{1}=0,\qquad b_{0}=-\frac{1+k}{2},
 \qquad R=-\frac{\cN^{\,2}-1}{24}(1+6k+k^{2}).
\end{align}
\emph{Supercharge:}
\begin{align}
E(q)=-\frac{\sn q(2\dn^{2}q-k^{\prime 2})}{\cn q\dn q},\qquad
 W(q)=-\frac{k(1+k)\sn^{2}q-k-1}{2\cn q\dn q}.
\end{align}
\emph{Potentials:}
\begin{align}
V_{\cN\!,1}^{\pm}=\frac{\cN^{\,2}k^{\prime 4}\sn^{2}q}{
 8\cn^{2}q\dn^{2}q}+\frac{\cN^{\,2}(1+k)^{2}}{8}
 \pm\frac{\cN k^{\prime 2}}{2}\biggl(\frac{k^{\prime 2}\sn q}{
 2\cn^{2}q\dn^{2}q}-\frac{k(1-k)\sn q}{2\dn^{2}q}\biggr).
\end{align}
\emph{Solvable sectors:}
\begin{subequations}
\begin{gather}
\cV_{\cN}^{(A)\pm}=(\cn q\dn q)^{\frac{\cN-1\pm 1}{2}}
 (1+k\sn q)^{\mp\frac{1}{2}}(1+\sn q)^{\mp\frac{1}{2}}
 \bra 1,\sn q,\dots,(\sn q)^{\cN-1}\ket, \\
\psi_{0}^{\pm}(q)\propto\biggl(\frac{\cn q\dn q}{
 (1+k\sn q)(1+\sn q)}\biggr)^{\mp\frac{\cN}{2}}.
\end{gather}
\end{subequations}

\subsection{Case V$'$: $A(z)=\frac{1}{2}(1-z^{2})(k^{\prime 2}
 -k^{2}z^{2})$, $z(q)=\cn q$}
\label{ssec:case5'}

\subsubsection{Case V$\,'\!$a: $b_{2}=b_{0}=0$}

\noindent
\emph{Parameters:}
\begin{align}
\label{eq:V'a0}
b_{2}=b_{0}=0,\qquad b_{1}=-\frac{1}{2},\qquad
 R=\frac{\cN^{\,2}-1}{12}(k^{\prime 2}-k^{2}).
\end{align}
\emph{Supercharge:}
\begin{align}
E(q)=\frac{\cn q(2\dn^{2}q-1)}{\sn q\dn q},\qquad
 W(q)=-\frac{\cn q}{2\sn q\dn q}.
\end{align}
\emph{Potentials:}
\begin{align}
V_{\cN\!,1}^{\pm}=\frac{\cN^{\,2}\cn^{2}q}{
 8\sn^{2}q\dn^{2}q}\pm\frac{\cN}{4}\biggl(
 \frac{1}{\sn^{2}q\dn^{2}q}-\frac{k^{2}-k^{\prime 2}}{
 \dn^{2}q}-1\biggr).
\end{align}
\emph{Solvable sectors:}
\begin{subequations}
\begin{gather}
\cV_{\cN}^{(A)\pm}=(\sn q)^{-\frac{\cN-1\pm 1}{2}}
 (\dn q)^{-\frac{\cN-1\mp 1}{2}}\bra 1,\cn q,\dots,(\cn q)^{\cN-1}
 \ket,\\
\psi_{0}^{\pm}(q)\propto\biggl(
 \frac{\sn q}{\dn q}\biggr)^{\mp\frac{\cN}{2}}.
\end{gather}
\end{subequations}

\subsubsection{Case V$\,'\!$b: $b_{1}=0$}

In this case, there are no real solutions of Eq.~\eqref{eq:cond8}
for $b_{2}$ and $b_{0}$.

\subsection{Case V$''$: $A(z)=\frac{1}{2}(1+z^{2})(1+k^{\prime 2}
 z^{2})$, $z(q)=\tn q$}
\label{ssec:case5''}

\subsubsection{Case V$\,''\!$a: $b_{2}=b_{0}=0$}

\noindent
\emph{Parameters:}
\begin{align}
b_{2}=b_{0}=0,\qquad b_{1}=\frac{k^{2}}{2},\qquad
 R=-\frac{\cN^{\,2}-1}{12}(1+k^{\prime 2}).
\end{align}
\emph{Supercharge:}
\begin{align}
E(q)=\frac{\sn q(k^{2}\cn^{2}q+2k^{\prime 2})}{\cn q\dn q},
 \qquad W(q)=-\frac{k^{2}\sn q\cn q}{2\dn q}.
\end{align}
\emph{Potentials:}
\begin{align}
V_{\cN\!,1}^{\pm}=-\frac{\cN^{\,2}}{8}\biggl(\dn^{2}q
 +\frac{k^{\prime 2}}{\dn^{2}q}-1-k^{\prime 2}\biggr)
 \mp\frac{\cN}{4}\biggl(\dn^{2}q-\frac{k^{\prime 2}}{
 \dn^{2}q}\biggr).
\end{align}
\emph{Solvable sectors:}
\begin{subequations}
\begin{gather}
\cV_{\cN}^{(A)\pm}=(\dn q)^{-\frac{\cN-1\mp 1}{2}}
 (\cn q)^{\cN-1}\bra 1,\tn q,\dots,(\tn q)^{\cN-1}\ket,\\
\psi_{0}^{\pm}(q)\propto (\dn q)^{\pm\frac{\cN}{2}}.
\end{gather}
\end{subequations}

\subsubsection{Case V$\,''\!$b: $b_{1}=0$}

In this case, there are no real solutions of Eq.~\eqref{eq:cond8}
for $b_{2}$ and $b_{0}$.

\section{Generalized Bender--Dunne Polynomials}
\label{sec:GBDP}

In this appendix, we summarize the generalized Bender--Dunne
polynomials (GBDPs) and present some explicit forms of them for
the shape-invariant and elliptic potentials investigated in
Sections \ref{sec:schro} and \ref{sec:ellip}. They are a natural
generalization of the ones first constructed for the quasi-exactly
solvable sextic anharmonic oscillator by Bender and Dunne
\cite{BD96} to all the type A $\cN$-fold supersymmetric models.
For more details, see Ref.~\cite{Ta03a}, Section~6.2. In
the unit system adopted in Sections \ref{sec:schro} and
\ref{sec:ellip}, the GBDPs $\pi_{n}^{[\cN]}(E)$ are characterized
by the following four-term recursion relation:
\begin{align}
\label{eq:4recu}
\pi_{n+1}^{[\cN]}(E)=&\,\bigl(E+2A_{n}^{[\cN]}\bigr)
 \pi_{n}^{[\cN]}(E)-4n(n-\cN)B_{n}^{[\cN]}\pi_{n-1}^{[\cN]}(E)
 \notag\\
&\,+8n(n-1)(n-\cN)(n-\cN-1)C_{n}^{[\cN]}\pi_{n-2}^{[\cN]}(E),
\end{align}
where $A_{n}^{[\cN]}$, $B_{n}^{[\cN]}$, and $C_{n}^{[\cN]}$
are given by
\begin{align}
A_{n}^{[\cN]}&=\biggl[n(n-\cN+1)+\frac{(\cN-1)(\cN-2)}{6}
 \biggr]a_{2}\mp\biggl(n-\frac{\cN-1}{2}\biggr)b_{1}+R,\\
B_{n}^{[\cN]}&=\left[\biggl(n-\frac{\cN}{2}\biggr)a_{1}
 \mp b_{0}\right]\left[\biggl(n-\frac{\cN}{2}\biggr)a_{3}
 \mp b_{2}\right],\\
C_{n}^{[\cN]}&=a_{4}\left[\biggl(n-\frac{\cN}{2}\biggr)a_{1}
 \mp b_{0}\right]\left[\biggl(n-\frac{\cN+2}{2}\biggr)a_{1}
 \mp b_{0}\right],
\end{align}
and the parameters $a_{i}$ and $b_{i}$ correspond to
the coefficients of the polynomials $A(z)$ in Eq.~\eqref{eq:defAz}
and $Q(z)$ in Eq.~\eqref{eq:defQz}, respectively. The degree
$\cN$ polynomial $\pi_{\cN}^{[\cN]}(E)$ in each polynomial
system $\{\pi_{n}^{[\cN]}(E)\}_{n=0}^{\infty}$ is called
$\cN$th critical GBDP. It was shown in Ref.~\cite{Ta03a} that
the anti-commutator of the type A $\cN$-fold supercharges is
(proportional to) the $\cN$th critical GBDP in the type A
Hamiltonian:
\begin{align}
\label{eq:antiA}
\bigl\{\bQ_{\cN}^{-},\bQ_{\cN}^{+}\bigr\}
 =\pi_{\cN}^{[\cN]}(\bH_{\cN}).
\end{align}

To obtain the recursion relation \eqref{eq:4recu} for a specific
case, we must first make a $GL(2,\bbC)$ transformation so that
the coefficient $a_{0}$ in $A(z)$ vanishes. For the elliptic
potential in Section~\ref{sec:ellip}, Case V$'$a, the canonical
form is $A(z)=\frac{1}{2}(1-z^{2})(k^{\prime 2}-k^{2}z^{2})$.
Thus, we transform $A(z)$ as
\begin{align}
\hA(z)=\Delta^{-2}(z+1)^{4}A\biggl(\frac{z-1}{z+1}\biggr)
 =\frac{1}{2}z\bigl[z^{2}+2(k^{\prime 2}-k^{2})z+1\bigr].
\end{align}
In our Case V$'$a, $Q(z)=-z/2$ from Eq.~\eqref{eq:V'a0} and it is
transformed simultaneously as
\begin{align}
\hQ(z)=\Delta^{-1}(z+1)^{2}Q\biggl(\frac{z-1}{z+1}\biggr)
 =-\frac{1}{4}(z^{2}-1).
\end{align}
Hence, the values of the parameters which should be substituted
into the recursion relation \eqref{eq:4recu} are as follows:
\begin{gather*}
a_{3}=\frac{1}{2},\qquad a_{2}=k^{\prime 2}-k^{2},\qquad
 a_{1}=\frac{1}{2},\qquad a_{4}=a_{0}=0,\\
b_{2}=-\frac{1}{4},\qquad b_{1}=0,\qquad
 b_{0}=\frac{1}{4},
\end{gather*}
and $R=(k^{\prime 2}-k^{2})(\cN^{2}-1)/12$ (cf. Eq.~\eqref{eq:V'a0}).
With these parameter values, the recursion relation \eqref{eq:4recu}
reduces to
\begin{align}
\pi_{n+1}^{[\cN]}(E)=\biggl[E+\frac{k^{\prime 2}-k^{2}}{2}
 (2n-\cN+1)^{2}\biggr]\pi_{n}^{[\cN]}(E)
 -n(n-\cN)\biggl(n-\frac{\cN+1}{2}\biggr)\biggl(
 n-\frac{\cN-1}{2}\biggr)\pi_{n-1}^{[\cN]}(E).
\end{align}
Using this three-term recursion relation, we can calculate, in
principle, the critical polynomial $\pi_{\cN}^{[\cN]}(E)$ for an
arbitrary value of $\cN$. For instance, we have
\begin{subequations}
\label{eqs:cgbdp}
\begin{align}
\pi_{1}^{[1]}(E)=&\,E,\\
\pi_{2}^{[2]}(E)=&\,(E-k^{2})(E+k^{\prime 2}),\\
\pi_{3}^{[3]}(E)=&\,E\bigl[E+2(k^{\prime 2}-k^{2})\bigr]^{2},\\
\pi_{4}^{[4]}(E)=&\,\bigl[E^{2}+2(2k^{\prime 2}-3k^{2})E+9k^{4}
 \bigr]\bigl[E^{2}+2(3k^{\prime 2}-2k^{2})E+9k^{\prime 4}\bigr],\\
\pi_{5}^{[5]}(E)=&\,E\bigl[E^{2}+10(k^{\prime 2}-k^{2})E
 +8(3k^{\prime 4}-2k^{\prime 2}k^{2}+3k^{4})\bigr]^{2},\\
\pi_{6}^{[6]}(E)=&\,\bigl[E^{3}+(16k^{\prime 2}-19k^{2})E^{2}
 +(64k^{\prime 4}-80k^{\prime 2}k^{2}+115k^{4})E-225k^{6}
 \bigr]\notag\\
&\,\times\bigl[E^{3}+(19k^{\prime 2}-16k^{2})E^{2}+(115k^{\prime 4}
 -80k^{\prime 2}k^{2}+64k^{4})E+225k^{\prime 6}\bigr],\\
\pi_{7}^{[7]}(E)=&\,E\bigl[E^{3}+28(k^{\prime 2}-k^{2})E^{2}
 +28(9k^{\prime 4}-10k^{\prime 2}k^{2}+9k^{4})E\notag\\
&\,+144(k^{\prime 2}-k^{2})(5k^{\prime 4}+2k^{\prime 2}k^{2}
 +5k^{4})\bigr]^{2},\\
\pi_{8}^{[8]}(E)=&\,\bigl[E^{4}+4(10k^{\prime 2}-11k^{2})E^{3}
 +2(264k^{\prime 4}-376k^{\prime 2}k^{2}+347k^{4})E^{2}\notag\\
&\,+4(576k^{\prime 6}-672k^{\prime 4}k^{2}+826k^{\prime 2}
 k^{4}-1155k^{6})E+11025k^{8}\bigr]\notag\\
&\,\times\bigl[E^{4}+4(11k^{\prime 2}-10k^{2})E^{3}
 +2(347k^{\prime 4}-376k^{\prime 2}k^{2}+264k^{4})E^{2}\notag\\
&\,+4(1155k^{\prime 6}-826k^{\prime 4}k^{2}+672k^{\prime 2}
 k^{4}-576k^{6})E+11025k^{\prime 8}\bigr],\\
\pi_{9}^{[9]}(E)=&\,E\bigl[E^{4}+60(k^{\prime 2}-k^{2})E^{3}
 +12(109k^{\prime 4}-146k^{\prime 2}k^{2}+109k^{4})E^{2}\notag\\
&\,+16(k^{\prime 2}-k^{2})(761k^{\prime 4}-118k^{\prime 2}k^{2}
 +761k^{4})E\notag\\
&\,+1152(35k^{\prime 8}-20k^{\prime 6}k^{2}+18k^{\prime 4}k^{4}
 -20k^{\prime 2}k^{6}+35k^{8})\bigr]^{2}.
\end{align}
\end{subequations}
In the limit $k\to 0$ ($k^{\prime}\to 1$), the trigonometric case
discussed in Sect.~\ref{sec:schro} is recovered, and each of
the above critical GBDP $\pi_{\cN}^{[\cN]}(E)$ exactly reduces to
the polynomial $\cP_{\cN}(E)$ in Eq.~\eqref{P(E)}. It is interesting
to note that all the above $\pi_{\cN}^{[\cN]}$ have the following
forms:
\begin{align}
\pi_{2M+1}^{[2M+1]}(E)
&=E\,\bar{\cR}_{M}^{(o)}(E;k^{2},k^{\prime 2})^{2},\\
\pi_{2M}^{[2M]}(E)&=\bar{\cR}_{M}^{(e)}(E;k^{2},k^{\prime 2})
 \bar{\cR}_{M}^{(e)}(E;-k^{\prime 2},-k^{2}),
\end{align}
where $\bar{\cR}_{M}^{(o)}$ and $\bar{\cR}_{M}^{(e)}$ are monic
polynomials of degree $M$ in the first variable, though we have not
been able to prove them for arbitrary $M$.

\bibliography{refsaps}
\bibliographystyle{npb}

\end{document}